\documentclass[graybox]{svmult}
\usepackage{cite}

\usepackage{mathptmx}       
\usepackage{helvet}         
\usepackage{courier}        
\usepackage{type1cm}        
\usepackage{epsfig,amsmath,amssymb,epstopdf}
\usepackage{makeidx}         
\usepackage{graphicx}        
\usepackage{multicol}        
\usepackage[bottom]{footmisc}

\usepackage[usenames,dvipsnames,svgnames,table]{xcolor}
\newcommand{\dt}{\Delta t}
\newcommand{\pd}{\partial}
\newcommand{\Gint}{G^{\rm int}}
\newcommand{\Gstokes}{G^{\rm Stokes}}
\newcommand{\bU}{\mathbf{U}}
\newcommand{\bu}{\mathbf{u}}
\newcommand{\nn}{\nonumber}
\usepackage[normalem]{ulem}

\makeindex             

\begin{document}

\title*{Order-of-magnitude speedup 
 for steady states and traveling waves via Stokes preconditioning 
  in Channelflow and Openpipeflow}
\titlerunning{Stokes preconditioning in Channelflow and Openpipeflow}
\author{Laurette S. Tuckerman and Jacob Langham and Ashley Willis}
\institute{Laurette S. Tuckerman \at Laboratoire de Physique et M\'ecanique des Milieux 
H\'et\'erog\`enes (PMMH),
CNRS, ESPCI Paris, PSL Research University; Sorbonne Universit\'e, 
Univ. Paris Diderot, France \\\email{laurette@pmmh.espci.fr}
\and Jacob Langham \at School of Mathematics, University of Bristol, 
Bristol, BS8 1TW, United Kingdom \\\email{J.Langham@bristol.ac.uk}
\and Ashley Willis \at School of Mathematics and Statistics, University of
Sheffield, Sheffield, S3 7RH United Kingdom \\\email{a.p.willis@sheffield.ac.uk}}
%
%
\maketitle

\abstract{
Steady states and traveling waves play a fundamental role in understanding
hydrodynamic problems. Even when unstable, these states provide the
bifurcation-theoretic explanation for the origin of the observed states. In
turbulent wall-bounded shear flows, these states have been hypothesized to be
saddle points organizing the trajectories within a chaotic attractor.  These
states must be computed with Newton's method or one of its
generalizations, since time-integration cannot converge to unstable equilibria.
The bottleneck is the solution of linear systems involving the Jacobian of the
Navier-Stokes or Boussinesq equations.  Originally such computations were
carried out by constructing and directly inverting the Jacobian,
but this is unfeasible for the matrices arising from three-dimensional
hydrodynamic configurations in large domains.  A popular method 
is to seek states that are
invariant under numerical time integration.  Surprisingly, 
equilibria may also be found by seeking flows that are invariant under 
a single very large Backwards-Euler Forwards-Euler timestep.  We show that this
method, called Stokes preconditioning, is 10 to 50 times faster at computing
steady states in plane Couette flow and traveling waves in pipe flow. Moreover,
it can be carried out using Channelflow (by Gibson) and Openpipeflow (by
Willis) without any changes to these popular spectral codes.  We explain the
convergence rate as a function of the integration period and Reynolds number by
computing the full spectra of the operators corresponding to the Jacobians of
both methods.}
%

\section{Motivation and history}
One of the fundamental properties of nonlinear systems is the existence 
  of multiple solutions. Many, in fact, most of these are
  unstable and hence not accessible to time integration. Yet, they play
  a crucial role in organizing the solution sets and possibly 
  the dynamics actually realized by the system.
  Bifurcation diagrams have long been studied in connection with classical
  hydrodynamic instabilities which form patterns,
  namely Taylor-Couette flow and Rayleigh-B\'enard convection.
  In 1989, a method was formulated by Tuckerman \cite{tuckerman1989steady}
  by which an explicit-implicit
  time integration code could be easily transformed to carry out
  steady state solving. More specifically, it was shown 
  that taking the difference between successive widely spaced timesteps
  was equivalent to preconditioning the Jacobian with the inverse
  Laplacian or Stokes operator.
  This allowed the linear systems whose solution is required by Newton's method
  to be solved economically by matrix-free iterative methods,
  which in turn allowed steady states in large systems ($O(10^5)$--$O(10^7)$
  degrees of freedom) to be computed.
  This method, called \emph{Stokes preconditioning},
  was applied to calculate bifurcation diagrams in
  spherical Couette flow by Mamun \& Tuckerman \cite{mamun1995asymmetry},
  convection in Cartesian \cite{xin1997natural,xin2002extended,xin2012stability,chenier1999stability,mercader2005bifurcations,henry2007multiple,dridi2008influence,torres2013three,torres2014bifurcation,torres2015transition}, cylindrical
  \cite{touihri1999onset,touihri2011instabilities,assemat2007nonlinear,marques2007centrifugal,boronska2010extreme,mercader2014secondary,sanchez2016natural} and spherical
  \cite{bergemann2008geoflow,feudel2011convection,feudel2013multistability,feudel2015bifurcations,feudel2017hysteresis} geometries by researchers such as Xin, Chenier, Henry, and Feudel, to von K\'arm\'an flow \cite{daube2002numerical,nore20031,nore2004survey}
  by Daube, Nore, and Le Qu\'er\'e 
  and to Bose-Einstein condensation \cite{huepe1999decay,abid2003gross} by Huepe, 
  Brachet and co-workers

  One of the problems to which Stokes preconditioning has been applied
  most extensively is that of double diffusive convection
  \cite{xin1998bifurcation,bergeon1998marangoni,bergeon1999double,bardan2000nonlinear,bergeon2002natural,bergeon2003soret,meca2004blue,meca2004complex,mercader2004numerical,batiste2004hydrodynamic,alonso2005numerical,alonso2007complex}.
  In 2006, a comprehensive theory of localized states via homoclinic
  snaking was developed by Knobloch and co-workers
  \cite{burke2006localized}, following earlier ideas
  by Champneys, Coullet, Fauve, Pomeau and others
  \cite{champneys1998homoclinic,coullet2000stable,fauve1990solitary,pomeau1986front,m1995pattern}.
  This theory predicts and explains the existence of a large number of
  branches linked by saddle-node bifurcations and thus spotlights the
  crucial role of unstable steady states and their computation.
  The first instances of localized states and
  homoclinic snaking in realistic physical systems were computed in 2006--8 by
  researchers such as Alonso, Assemat, Batiste, Bergeon, and Mercader 
  \cite{batiste2006spatially,alonso2007numerical,bergeon2008spatially,bergeon2008periodic,assemat2008spatially},
  all for double diffusive convection and all using the Stokes
  preconditioning method, followed by further studies of these
  phenomena by Beaume, LoJacono and others
  \cite{lo2010spatially,beaume2011homoclinic,mercader2011convectons,beaume2013convectons,beaume2013nonsnaking,mercader2013travelling,jacono2017spatially,jacono2017localized,jacono2017complex}.
  In the Appendix, we will show examples of the more exotic or
  interesting bifurcation diagrams that have been calculated by
  various groups using the Stokes preconditioning method.

  The cases studied above involved flows which undergo linear instabilities
  from a homogeneous state, leading to patterns.
  We now turn to shear flows confined between rigid or free-slip boundaries,
  namely plane Couette flow, plane Poiseuille flow and pipe flow, 
  which are linearly stable over the parameter range of interest.
  In 1990, Nagata \cite{Nagata1990three} computed the first non-trivial steady state
  for plane Couette flow; this was followed by computations of other
  steady states and traveling wave solutions for plane Couette and Poiseuille flow
  by Waleffe \cite{waleffe1998three,waleffe2001exact,waleffe2003homotopy}
  who coined the term \emph{Exact Coherent Structure} (ECS) to describe
  unstable steady states, traveling waves, periodic orbits, and other
  low-dimensional dynamically relevant invariant solutions with low-dimensional
  unstable manifolds in these shear flows.
  These and other contemporaneous computations
  \cite{clever1997tertiary,nagata1997three,faisst2000transition,Schmiegel_PhD}
  used dedicated steady-state codes in which the Jacobians were constructed
  and the linear systems required by Newton's method were solved directly.
  Many used homotopy to modify steady states or traveling waves already
  known to exist in convection or in Taylor-Couette flow.
  In 2003--4, the first traveling waves in pipe flow were computed 
  by Eckhardt, Kerswell and co-workers \cite{faisst2003traveling,wedin2004exact},
  also using Jacobian matrices and direct matrix inversion.
  All of these solutions were computed in small periodic domains or 
  \emph{minimal flow units}, which are the smallest domains in which turbulence
  can be maintained.
  The motivation for these searches was provided by Cvitanovi{\'c} and Eckhardt
  \cite{cvitanovic1991periodic,Cvitanovic_Eckhardt}
  and by Kawahara \cite{kawahara2001periodic,kawahara2012significance}.
  Building on previous ideas by Smale, Ruelle, Bowen and Sinai, the authors
  proposed that turbulence could be viewed from a deterministic dynamical systems
  perspective as a collection of trajectories ricocheting between ECSs along
  their unstable directions.

  Influential work on numerical methods
  \cite{dennis1996numerical,knoll2004jacobian,sanchez2004newton,viswanath2007recurrent,van2011matrix}
  by authors such as S\'anchez, Net, van Veen, Viswanath, and others
  led to the adoption of iterative matrix-free methods and to many additional solutions
  \cite{pringle2007asymmetric,kerswell2007recurrence,
    duguet2008transition,duguet2008relative,gibson2008visualizing} by reseachers
  such as Duguet, Gibson, Halcrow, Pringle and Willis.
  Catalogs of solutions computed by 2009 were given in
  \cite{Schmiegel_PhD,gibson2009equilibrium,pringle2009highly}. 
  In a development parallel to what was happening at the same time 
  in double diffusive convection, focus was extended from the minimal flow unit
  to localized and snaking solutions. Since these consist of one or more
  active regions surrounded by possibly wide quiescent regions, 
  computations of localized states are far more costly.
  The codes Channelflow \cite{gibson2008visualizing,channelflow} by Gibson
  and Openpipeflow \cite{willis2017openpipeflow} by Willis
  took advantage of the matrix-free iterative approach to treat larger domains, 
  leading to the discovery of localized states in plane Couette flow by
  Schneider et al.~\cite{schneider2010snakes} in 2010 and in pipe flow by 
  Avila et al.~\cite{avila2013streamwise} in 2013,
  followed by many others 
  \cite{gibson2014spanwise,brand2014doubly,eckhardt2014doubly,chantry2014genesis}. 
Mellibovsky and Eckhardt \cite{mellibovsky2011takens} used Stokes preconditioning to investigate a Takens-Bogdanov scenario for travelling waves in pipe flow. Beaume and co-workers 
  \cite{beaume2015reduced,beaume2016modulated} used the
  method to calculate ECSs for an asymptotic reduction
  of the free-slip version of plane Couette flow called Waleffe flow
  \cite{waleffe1998three,waleffe2001exact,waleffe2003homotopy} and
  were able to reproduce the high-Reynolds-number scaling calculated
  in \cite{wang2007lower,blackburn2013lower,deguchi2014high}.


  The primary purpose of this Chapter is to compare the method
  used in Channelflow and Openpipeflow with the Stokes preconditioning method.
  We show that the Stokes preconditioning method can be carried out
  by using the options already present in Channelflow and Openpipeflow without 
  any changes to the codes themselves. The user must simply liberate him or herself
  from standard notational convention and allow the timestep $\dt$ to approach infinity, 
  as will be explained in Section \ref{sec:stokes}. In Section \ref{sec:channelpipe},
  we will give examples of steady states and traveling waves calculated by
  Channelflow and Openpipeflow using the Stokes preconditioning method. 
  We will show that these computations are up to 50 times
  faster than when they are carried out via the classical approach previously used by
  Channelflow and Openpipeflow, emphasizing again that this is accomplished
  only by the choice of different parameters.

\section{Stokes preconditioning}
\label{sec:stokes}
We begin with the schematically written differential equation, or dynamical system:
\begin{equation}
  \frac{\pd U}{ \pd t}=LU + N(U),
\label{eq:DE}\end{equation}
for a flow-field $U$, where $L$ stands for the linear terms, typically the viscous or diffusive terms
in hydrodynamics and $N$ for the nonlinear terms, typically the advective terms.
We seek steady states of \eqref{eq:DE}, i.e. roots of $L+N$, or $U$ such that
\begin{equation}
0=LU + N(U).
\label{eq:func1}\end{equation}
There are other ways to define steady states. In particular, we may seek
\begin{equation}
0 = U(T)-U(0) \equiv (\Phi_T - I)U(0),
\label{eq:func2} \end{equation}
where $U(T)$ is computed from $U(0)$ by time-integration of \eqref{eq:DE}, and
where $\Phi_T$ is the operator which takes an initial condition $U(0)$ to the
field $U(T)$ at time $T$. We note that, in addition to equilibria,
definition \eqref{eq:func2} is satisfied by $T$-periodic orbits 
of the dynamical system, which we will not consider here.

In practice, temporal integration of nonlinear evolution equations
cannot be carried out exactly. A notable exception to this,
useful for conceptual purposes, is if $N$ is in fact linear, so that
we  may write
\begin{equation}
  U(T)= e^{(L+N)T} U(0).
  \label{eq:exp}  \end{equation}
In general, however, $\Phi_T$ must be approximated as the product of many
small approximate timesteps.
That is,
  \begin{equation}
    \Phi_T \approx \left(B_{\dt}\right)^{T/\dt},
\label{eq:littlesteps}  \end{equation}
where $B_{\Delta t}$ is a numerical timestepping operator with step size
$\Delta t$.
One straightforward approach combines backward-Euler timestepping
for $L$ with forward-Euler timestepping for $N$, leading to the BEFE algorithm
  \begin{equation}
	  B_{\dt} \equiv (I-\dt L)^{-1}(I+\dt N),
\label{eq:befe}  \end{equation}
which is used in the Stokes preconditioning method.

Despite the use of the notation $L$ and $N$,
the necessary distinction is actually between the implicitly and explicitly
integrated parts of the operator; although $L$ must be linear, $N$ may include
some of the linear terms.
In the interests of keeping the notation simple, 
we will not always be rigorous about distinguishing between
an operator such as $N$ and its linearization $N_U$ about a solution $U$. 
We will also not distinguish between the spatially continuous equations and
their spatially discretized versions. Nor will we consider the
constraints, i.e. boundary conditions, incompressibility, and the
crucial related question of determination of the pressure,
assuming the constraints to be incorporated into the
implementation of $(I-\dt L)^{-1}$,
although we will later briefly address these questions.

We make a few comments about \eqref{eq:befe}.
First, the expression $(I-\dt L)^{-1}$ should be understood
as making a rational approximation to $e^{\dt L}$, 
called \emph{implicit}, meaning in this case that it
requires an operator inversion.
The expression $(I+\dt N)$ should be understood as a polynomial
approximation to $e^{\dt N}$ and is called \emph{explicit} because it 
requires only operator actions.
The reason for which $L$ is integrated implicitly is that it has widely
spaced eigenvalues, a property called \emph{stiffness} in the context of differential
equations. Stiffness implies that a polynomial (explicit) approximation of the
exponential would require unacceptably small $\dt$ in order to remain finite.
The polynomial approximation $(I + \dt N)$ also limits the size of $\dt$,
but far less than the limitation that would be imposed by the explicit
integration of $L$.
It is very fortunate that it is the linear term $L$ that bears most of
the responsibility for the stiffness of \eqref{eq:DE}, 
since implicit formulas involving the nonlinear term $N$ would
be much harder to evaluate, requiring Newton iteration at each timestep,
rather than merely the inversion of a linear operator.
Hence, mixed explicit-implicit formulas such as \eqref{eq:befe} are almost
universally used in hydrodynamics.

For small $\dt$, the Taylor expansion of \eqref{eq:befe} shows that 
\begin{eqnarray}
  B_{\dt} &\equiv& (I-\dt L)^{-1}(I+\dt N) \nn\\
          &\approx&   (I + \dt L + (\dt L)^2 + \ldots)(I + \dt N) \nn\\
&\approx& I + \dt (L+N) + \dt^2 (L^2 + LN) + \ldots 
\label{eq:Taylorbefe}\end{eqnarray}
while 
\eqref{eq:exp} shows that the exact flow satisfies
\begin{equation}
	\Phi_{\dt} = I + \dt (L+N) + \dt^2 (L+N)^2/2 + \ldots
\label{eq:Taylorflow}\end{equation}
Thus, \eqref{eq:befe} is first-order accurate, meaning that
the Taylor series agree to order $\dt$.
Note that the disagreement between \eqref{eq:Taylorbefe}
and \eqref{eq:Taylorflow}, here at order $\dt^2$,
is not a function of $L+N$, a property called \emph{time-splitting error}.
Other, more accurate, timestepping operators may be used
instead of BEFE to approximate $\Phi_T$, but 
time-splitting errors are always present
when different formulas are used to integrate $L$ and $N$.

Let us now calculate the increment produced by evolving with
the Backwards-Euler Forwards-Euler method \eqref{eq:befe}
over a single timestep:
\begin{svgraybox}
\begin{eqnarray}
  B_{\dt}-I &=& (I-\dt L)^{-1}(I+\dt N) - I  \nn \\
  &=& (I-\dt L)^{-1}\left[(I+\dt N) - (I-\dt L)\right] \nn \\
          &=& (I-\dt L)^{-1}\dt(L+N).
              \label{eq:mycalc}\end{eqnarray}
\end{svgraybox}
\noindent We emphasize that, unlike \eqref{eq:Taylorbefe}, equation
\eqref{eq:mycalc} is an exact algebraic
calculation and not a Taylor expansion: it is valid for all values of $\dt$.
Thus, \eqref{eq:mycalc} demonstrates the
simple and powerful result that roots of
$B_{\dt}-I$ are also exact roots of $L+N$, regardless of the value of $\dt$.
Equation \eqref{eq:mycalc} follows from the definition \eqref{eq:befe}
of the BEFE algorithm: it is not a general property of timestepping schemes.

We recall that a wide distribution of eigenvalues of an operator
is called stiffness in the context of differential equations; 
the same property is called \emph{poor conditioning} in the context of
linear operators. Poor conditioning can be counteracted by
preconditioning, i.e. multiplication by an operator which resembles
the inverse of the poorly conditioned operator,
but whose action is more easily or cheaply carried out.
Thus, \eqref{eq:mycalc} shows that $B_{\dt}-I$ is 
a version of $L+N$ that has been preconditioned by $(I-\dt L)^{-1}$.
Since, in the hydrodynamic context, $\pd U/\pd t=LU$ 
is the time-dependent Stokes (rather than Navier-Stokes) equation,
for which $(I-\dt L)^{-1}$ is the (implicit) timestepping operator,
we call this \emph{Stokes preconditioning}.

In the limit of small $\dt$, we have 
\begin{equation}
  (B_{\dt}-I) = (I-\dt L)^{-1}\dt(L+N ) \approx \dt (L+N),
\label{eq:smalldt}\end{equation}
while in the limit of large $\dt$, we have
\begin{equation}
  (B_{\dt}-I) =   (I- \dt L)^{-1} \dt (L+N) \approx -L^{-1}(L+N).
\label{eq:largedt}\end{equation}
Varying $\dt$ interpolates between these two cases, that
of $L+N$ preconditioned by $L^{-1}$ and that of $L+N$ itself.
(The conditioning or stiffness of operators is unaffected
by scalar multiplication, since it is the ratio between
eigenvalues or timescales that is significant.)
We emphasize that Stokes preconditioning is not carried out as a separate
operation:
that is, we do not apply $L+N$ followed by $(I-\dt L)^{-1}$. Instead,
we carry out $B_{\dt}-I$, which \eqref{eq:mycalc} shows to be 
equivalent to the product of these two operators.
In what follows, we will write $B_T-I$, keeping in mind that $T$ may be of any size. 

We summarize what we have said above by defining various nonlinear operators $G$
whose roots are the exact, or approximate, steady states 
of \eqref{eq:DE}. We define two theoretical operators:
\begin{eqnarray}
	G^{\rm rhs} &\equiv& L+N, \label{eq:rhs}\\
	G^{\rm flow} = G^{\rm flow}_T &\equiv& \Phi_T-I, \label{eq:flow}
\end{eqnarray}
whose roots are steady states of the underlying dynamical system~\eqref{eq:DE}, in
accordance with the definitions given in~\eqref{eq:func1} and~\eqref{eq:func2}
respectively.
These roots may be computed by seeking the zeros of the following two numerical
operators,
whose comparison will be the main focus of this Chapter:
\begin{svgraybox}
\begin{eqnarray}
  \Gstokes = \Gstokes_T &\equiv& B_T-I,\label{eq:Gstokes}\\
  \Gint = \Gint_T &\equiv&
	\left(B_{\dt}\right)^{T/\dt}-I. \label{eq:Gint}
\end{eqnarray}
\end{svgraybox}
\noindent We omit the time $T$ when it is not important.

These two operators are fundamentally different.
Unlike $\Gint$, $\Gstokes$ is not a time-discretized version of $G^{\rm flow}$.
Instead, as shown in \eqref{eq:mycalc}, the roots of $\Gstokes$ are those of
$G^{\rm rhs}$ if BEFE is used for $B_T$.
This is true for any value of $T$, but only a large value of $T$ leads to a 
well-conditioned Jacobian.
In contrast, $\Gint$ corresponds to standard time-integration of $G^{\rm flow}$.
Although any scheme including BEFE may be used, typically higher-order methods,
such as Crank-Nicolson Adams-Bashforth or semi-implicit Runge-Kutta are preferred.
For $\Gint$ to approximate $G^{\rm flow}$, $\dt$ must have a small value
in order for the time integration to be accurate and stable.
With the usual non-dimensionalization of space and time,
and at Reynolds or Rayleigh numbers in the pattern-forming or transitional
range, this usually means $\dt = O(0.01)$. 
Despite their very different natures, the coding for the two operators
is identical. 
The difference is only quantitative: for $\Gstokes$, one BEFE timestep is taken,
with $\dt=T$ large, while for $\Gint$, many timesteps with small $\dt$ are taken.

Steady states, i.e. roots of an operator $G$, are calculated via Newton's method.
For an estimated steady state $U$, we write the linear approximation to $G$
\begin{equation}
  G(U -u) \approx G(U) - G_U u,
  \label{eq:linear}\end{equation}
where $G_U$ is the linearization of $G$ about $U$ and select $u$ such as to make
the right-hand-side of \eqref{eq:linear} zero:
\begin{equation}
  G_U u = G(U),
\label{eq:biglin}\end{equation}
updating $U$ as
\begin{equation}
U \leftarrow U-u.
\end{equation}

Traveling waves can be considered to be steady states in a moving frame.
Rather than $\pd U/\pd t=0$, waves traveling in the $x$-direction satisfy
\begin{equation}
\frac{\pd U}{\pd t} = -C \frac{\pd U}{\pd x},
\end{equation}
where the wavespeed $C$ is an additional unknown.
Therefore, they may be computed as steady states of 
\begin{equation}
0 = LU + N(U) +C \frac{ \pd U}{ \pd x},
\end{equation}
by adding $C \pd U/\pd x$ to the operator $N$
and imposing an additional condition to fix the spatio-temporal
phase of the traveling wave.

The bottleneck for Newton's method is the solution of the linear system
\eqref{eq:biglin}.  Assuming that the dimension of $G_U$ is too large for
\eqref{eq:biglin} to be solved directly via Gaussian elimination, it must be
solved iteratively using a matrix-free approach that avoids explicit
construction of the Jacobian.  The method of choice is conjugate gradient
iteration or, rather, one of its suitable generalizations for non-symmetric
positive definite operators $G_U$.
The most widely used of these is GMRES, Generalized Minimum RESidual \cite{saad1986gmres}, 
but BiCGSTAB \cite{van1992bi} also has its adherents.
These methods calculate an approximation to $u$ in a Krylov space, i.e.
in the space spanned by successive actions of the linear operator $G_U$ on
$G(U)$:
\begin{equation}
  G(U),\; G_U \,G(U),\; G_U^2 \,G(U),\; \ldots \; G_U^k \, G(U),\; \ldots
  \;G_U^{K-1} \,G(U).
\label{eq:Kspace}\end{equation}
The CPU time required to solve \eqref{eq:biglin} thus depends on two factors:
\vspace*{-0.3cm}
\begin{eqnarray}
  && \mbox{--the time required for a single action by $G_U$},
     \qquad\qquad\qquad\qquad\qquad\qquad\qquad \nn\\[-0.1cm]
&&\mbox{~~~~determined by the number of timesteps $T/\dt$, and} \label{eq:factor1}\\
&& \mbox{--the number $K$ of actions needed to approximate the solution $u$,}\nn\\[-0.1cm]
&&   \mbox{~~~~determined by the conditioning of $G_U$.}
  \label{eq:factor2}
\end{eqnarray}
\vspace*{-0.2cm}
The action of $G_U$ on $u$ can be carried out by taking $N\rightarrow N_U$,
i.e. substituting
\begin{equation}
  (\bU\cdot\nabla)\bU \rightarrow (\bu\cdot\nabla)\bU + (\bU\cdot\nabla)\bu
\end{equation}
and replacing all inhomogeneous terms such as boundary conditions,
pressure gradients or fluxes, by corresponding homogeneous terms.
This can typically be accomplished with very little modification to a code.
Alternatively, it can be approximated via finite differences as 
\begin{equation}
  [G(U+\epsilon u) - G(U)]/\epsilon
\label{eq:finitediff}\end{equation}
In either case, evaluation of $G_U u$ typically takes
approximately the same time as evaluation of $G(U)$.
Therefore the time required to compute one such action is proportional 
to the number of timesteps $T/\Delta t$.
For an optimized pseudospectral code,
the most time-consuming portion of a timestep is spent in the
spectral-to-physical transforms, which take a time roughly proportional to
\begin{equation}
M_x M_y M_z\left(\log M_x +\log M_y + \log M_z\right) \equiv M\log M ,
\end{equation}
where $M_x, M_y, M_z$ are the number of gridpoints or modes
in each of the three Cartesian directions and $M$ is their product.
For finite-difference discretizations, 
the most time-consuming portions may be the inversion of elliptic operators,
which in tensor-product geometries can be carried out
\cite{tuckerman1989steady,lynch1964direct,haidvogel1979accurate,vitoshkin2013direct}
in a time proportional to
\begin{equation}
M_x M_y M_z\left(M_x + M_y + M_z\right)\sim M^{4/3}.
\end{equation}
We will consider that the CPU time spent in carrying out actions by $G_U$
is proportional to $K\; (T/\dt) \; M^{1+\alpha}$ with $0\leq \alpha \leq 1$.
The number $K$ of actions is determined by the nature of the operator $G_U$,
more specifically its spectrum, a simple measure of which is given by its 
condition number, which is roughly the ratio of absolute values of the
largest to the smallest eigenvalue.
The total number of timesteps taken in each Newton step is the product
$K \times T/\dt$.

Other considerations can enter into the timing.
The most important one we have found is related to the GMRES algorithm
without restarts, which requires the orthogonalization of all of 
the Krylov vectors \eqref{eq:Kspace} to each other, requiring
$K(K-1)/2$ scalar products of vectors of length proportional to $M$.
For $K$ sufficiently large, the CPU time required to take these scalar
products may approach or even exceed that taken by the timesteps.
We may write the total time required to take one Newton step as 
\begin{equation}
  {\rm CPU} \approx c_{\rm action} \frac{T}{\dt} K M^{1+\alpha}  + c_{\rm
  orth} K^2 M,
\label{eq:CPU}\end{equation}
where $c_{\rm action} \gg c_{\rm orth}$ are prefactors measuring the
CPU time taken to perform a timestep or a scalar product.


\section{Stokes preconditioning using Channelflow and Openpipeflow}
\label{sec:channelpipe}

We have carried out computations of steady states in
plane Couette flow using Channelflow and traveling waves 
in pipe flow using Openpipeflow.
In the plane Couette case, the computations using the Stokes preconditioning method
take about 10\% of the time required by computations using the classic integration method when measured in terms of number of timesteps required. 
For pipe flow, the Stokes preconditioning method is 35 to 50 times faster
than the classic integration method.
This section gives details of these computations and explores the
conditioning properties of the operators $\Gstokes$ and $\Gint$.

\subsection{The flows}
The mathematical framework for plane channel flows assumes
flow between infinite parallel plates,
with velocities which are prescribed on the plates
and which are assumed periodic in the two directions parallel to the plates.
Each periodic boundary condition must be completed by specifying
either the flux or the pressure gradient in that direction.
Plane Couette flow prescribes a finite velocity difference
between the plates and no pressure gradient in the directions parallel to the plates,
while standard plane Poiseuille flow prescribes no velocity difference
between the plates, a finite pressure gradient or flux in one
parallel direction and zero flux in the other.
The direction between the plates is called \emph{cross-channel} or
\emph{wall-normal}, 
the direction of the fixed velocity difference in the Couette case or
of the fixed pressure gradient or flux in the Poiseuille case is called
\emph{streamwise} and the remaining direction is called \emph{spanwise}.
(Any combination of Poiseuille and Couette flow is possible and,
in addition, the flux can be chosen to be zero in either flow
by the appropriate choice of velocity at the walls.)
Pipe flow, in a cylindrical geometry, also has two periodic directions,
with a prescribed velocity on the pipe surface, 
a pressure gradient or flux in the streamwise
direction and zero pressure gradient in the azimuthal direction.

Although the three flows are qualitatively different, their
overall behavior is extremely similar.
All have simple laminar states which are homogeneous in the periodic directions,
All undergo transition to turbulence despite the fact that 
the laminar states are linearly stable in the Reynolds number range relevant for transition.
(Plane Couette flow and pipe flow are linearly stable for all Reynolds numbers,
while plane Poiseuille flow becomes unstable for a Reynolds number which is
considerably higher than that at which transition takes place.)
All possess an abundance of unstable states, which we investigate below.
Two differences are relevant for our discussion.
First, when there is an imposed flux (or pressure gradient)
along the streamwise direction of the pipe or channel,
states which break the streamwise homogeneity are not steady, 
but traveling waves or more complex flows.
Second, the Reynolds numbers defined for the three flows 
use different conventions.
Since for all of these flows the instability mechanism is shear, 
a common Reynolds number based on a single-sign shear layer
would be more appropriate. Laminar plane Couette flow
($U_{\rm lam}=y$ over $-1\leq y \leq 1$)
possesses one single-sign shear layer
with $\Delta U = 2$ and $\Delta Y = 2$, while
the $y=0$ plane divides laminar plane Poiseuille flow
($U_{\rm lam}=3(1-y^2)/2$ over $-1\leq y \leq 1$)
into two opposite-sign shear layers 
each with $|\Delta U|=3/2$ and $\Delta Y = 1$, which yields a factor of
$4/(3/2) = 8/3=2.67$ between the conventional Reynolds numbers for
Couette and Poiseuille flow.
For pipe flow ($U_{\rm lam}=2(1-4r^2)$ over $0 \leq r \leq 1/2$)
the length scale used in defining the Reynolds number
is the diameter, so that $\Delta U=2$ and $\Delta Y = 1/2$.
Hence the conventional Reynolds numbers for pipe flow
are roughly four times those for plane Couette flow with similar phenomenology.
Evidence for this is that transition to turbulence begins in
plane Couette flow at $Re\sim 325$ \cite{shi2013scale},
in plane Poiseuille flow at $Re\sim 1000$,
and in pipe flow at $Re\sim 2040$ \cite{avila2011onset},
a factor of 6.3 higher than for transition in plane Couette flow.
The lowest known non-trivial states appear at 127 for plane Couette flow
\cite{clever1997tertiary,waleffe2003homotopy} and at 773 for pipe flow \cite{pringle2009highly},
a factor of 6.

\subsection{The codes}

The Channelflow \cite{gibson2008visualizing,channelflow}
and Openpipeflow \cite{willis2017openpipeflow} software packages
are widely used in the hydrodynamic stability community.
As their names imply, Channelflow and Openpipeflow
simulate flows in a planar and in a cylindrical geometry, respectively.
Both treat tensor-product geometries, with two periodic directions
and one non-periodic direction. 
Both represent the two periodic directions as Fourier series,
taking derivatives in the Fourier space, where they are sparse.
This means that inverting Laplacian ($L$) and Helmholtz ($I-\dt L$)
operators is cheap and easy: these inversions are carried
out exactly and not iteratively, contrary to what is done in
spectral element or finite volume codes meant to treat general geometries.
Both use the influence matrix technique to impose incompressibility
so that the solutions calculated are divergence-free to machine
accuracy and not merely to some power of $\dt$.
In addition to timestepping, i.e. action by 
$B_{\dt}$, both have also implemented steady-state solving,
using Newton's method to find roots of $\Gint$.
It can be seen from \eqref{eq:Gstokes} and \eqref{eq:Gint} that 
the operator $\Gstokes$ can be represented via $\Gint$ 
merely by setting the parameters $T$ and $\dt$ to be equal and selecting
the BEFE scheme to perform the timestepping, rather than the higher
order methods that are usually preferred for evaluation of $\Gint$.
No changes at all need to be made to the codes, which 
makes them ideal for comparing 
the calculation of steady states and traveling waves
via $\Gint$ versus $\Gstokes$.

We complete the description of the two codes by mentioning
a few other elements which are less relevant for our purpose.
Channelflow discretizes the spatial derivatives in the direction
between the two bounding planes via
Chebyshev polynomials, while Openpipeflow discretizes the
nonperiodic radial direction via finite differences.
Both codes use a Chebyshev grid spacing in the nonperiodic direction,
with points concentrated near the solid boundaries.
Both use the pseudospectral method, with derivatives taken
in the Fourier space and multiplications carried out by
transforming to a gridspace representation, multiplying the
values at each point and then transforming back.
Both use GMRES to solve the linear systems \eqref{eq:biglin}.
Both approximate the action of $G_U$ as a difference
\eqref{eq:finitediff} between two actions of $G$.
Both implementations improve the convergence radius of Newton's method
by including the complementary hookstep \cite{viswanath2007recurrent}
feature, which constrains the Newton step
so that it remains within a trusted region of validity for
linearization \eqref{eq:linear}.

\subsection{Computations with Channelflow}

\begin{figure}
\includegraphics[width=\textwidth]{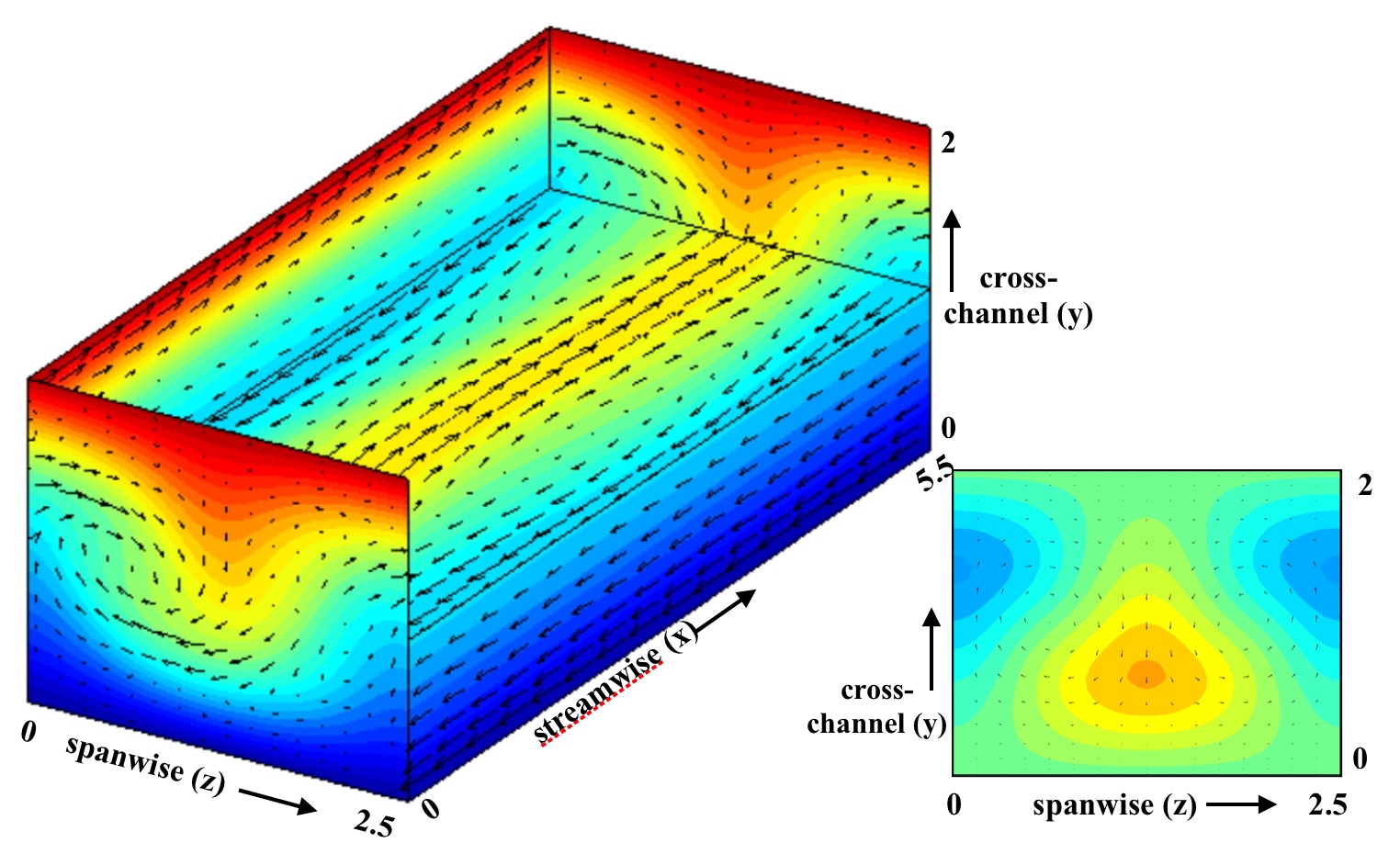}
\caption{Visualization of the steady state EQ1, also called NCBW.
    Left: Cutaway perspective view. Colors show streamwise velocity, arrows show velocity
  in various planes. 
  Right: Streamwise cross-section showing deviation of streamwise velocity
  from the laminar state.
Adapted from Gibson, Halcrow, Cvitanovi\'c \cite{gibson2009equilibrium}.}
\label{fig:EQ1NCBW}
\end{figure}
\begin{figure}
\includegraphics[width=0.5\textwidth]{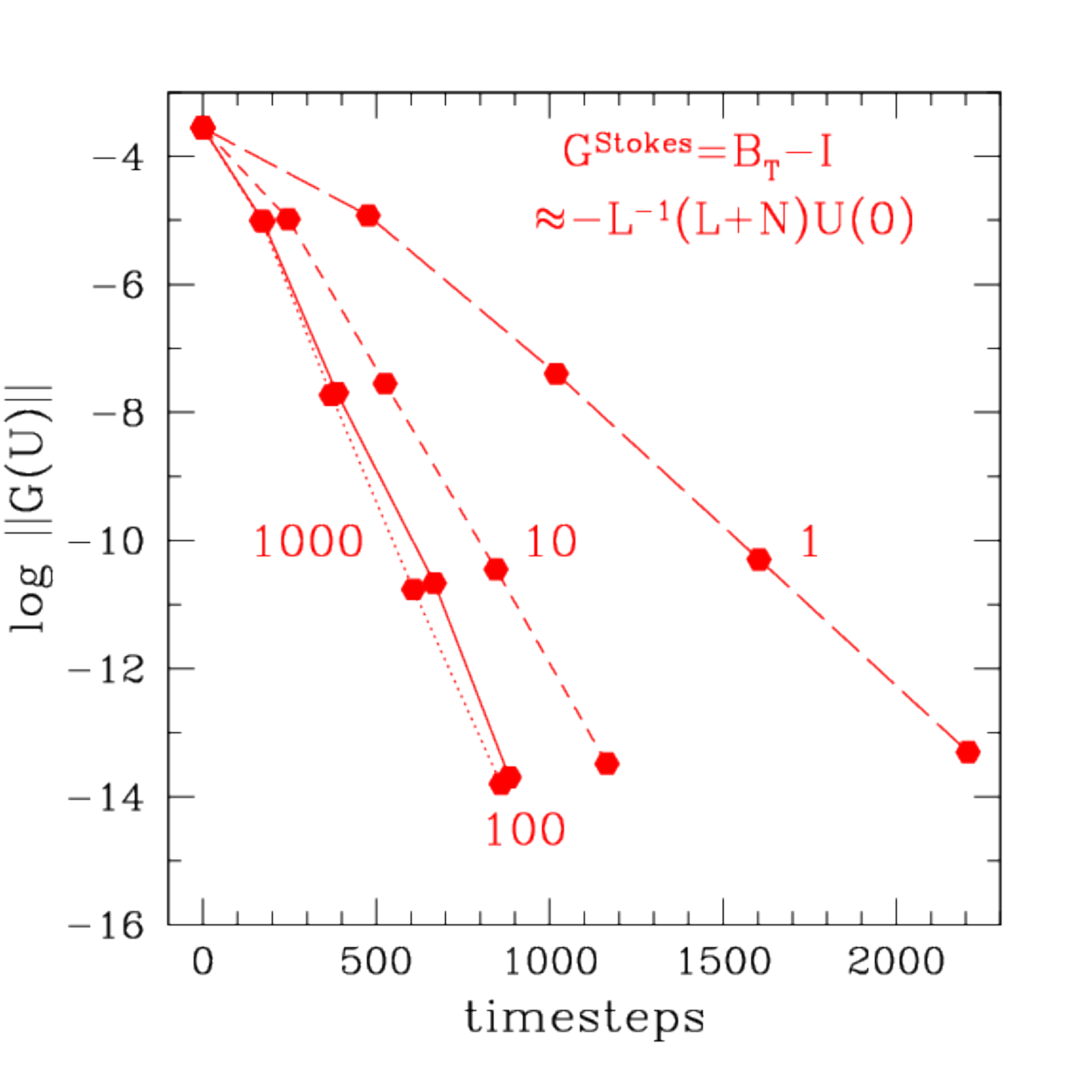}
\includegraphics[width=0.5\textwidth]{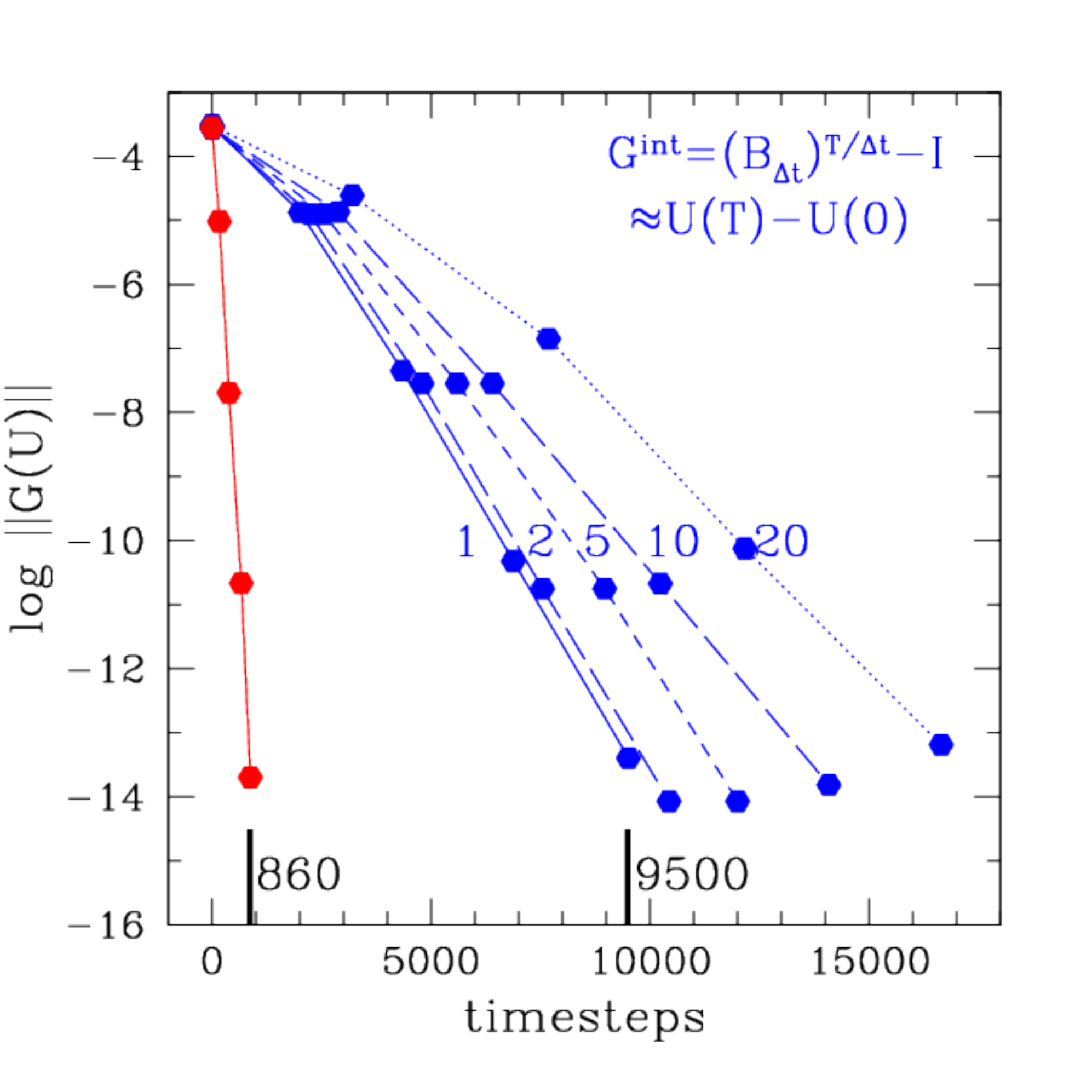}
\caption{Performance of Newton's method with GMRES in Channelflow for
  single-step Stokes preconditioning method $\Gstokes_T$ and for
  multi-step integration method $\Gint_T$ for various values of $T$.
  The initial condition is the steady EQ1 state \cite{Nagata1990three,clever1997tertiary,waleffe1998three,waleffe2003homotopy} at $Re=400$ and $Re$ is lowered to 380.
  Ordinate shows $\log||G(U)||$ at each Newton iteration, while abscissa shows
  the number of timesteps taken thus far. Each curve is labelled by its value of $T$.
  Left: Performance of single-step $\Gstokes_T \equiv B_T-I$.
  Convergence is fastest when $T$ is highest, achieving
  approximately asymptotic performance for $T=100$ (solid curve).
  Right: Performance of multi-step $\Gint_T \equiv (B_{\dt})^{T/\dt}-I$
  with timestep $\dt=0.031\,25$.
  Convergence is fastest when $T$ is lowest, achieving approximately
  asymptotic performance for $T=1$ (solid curve). Shown for comparison is the
  convergence curve for $\Gstokes$ with $T=100$.
  Short black lines highlight the difference in speed between the two methods:
  the classic integration method takes 9500 timesteps in this case, while 
  the Stokes method takes 860 timesteps, a ratio of 11.}
\label{fig:classic_and_stokes}
%
\includegraphics[width=0.5\textwidth]{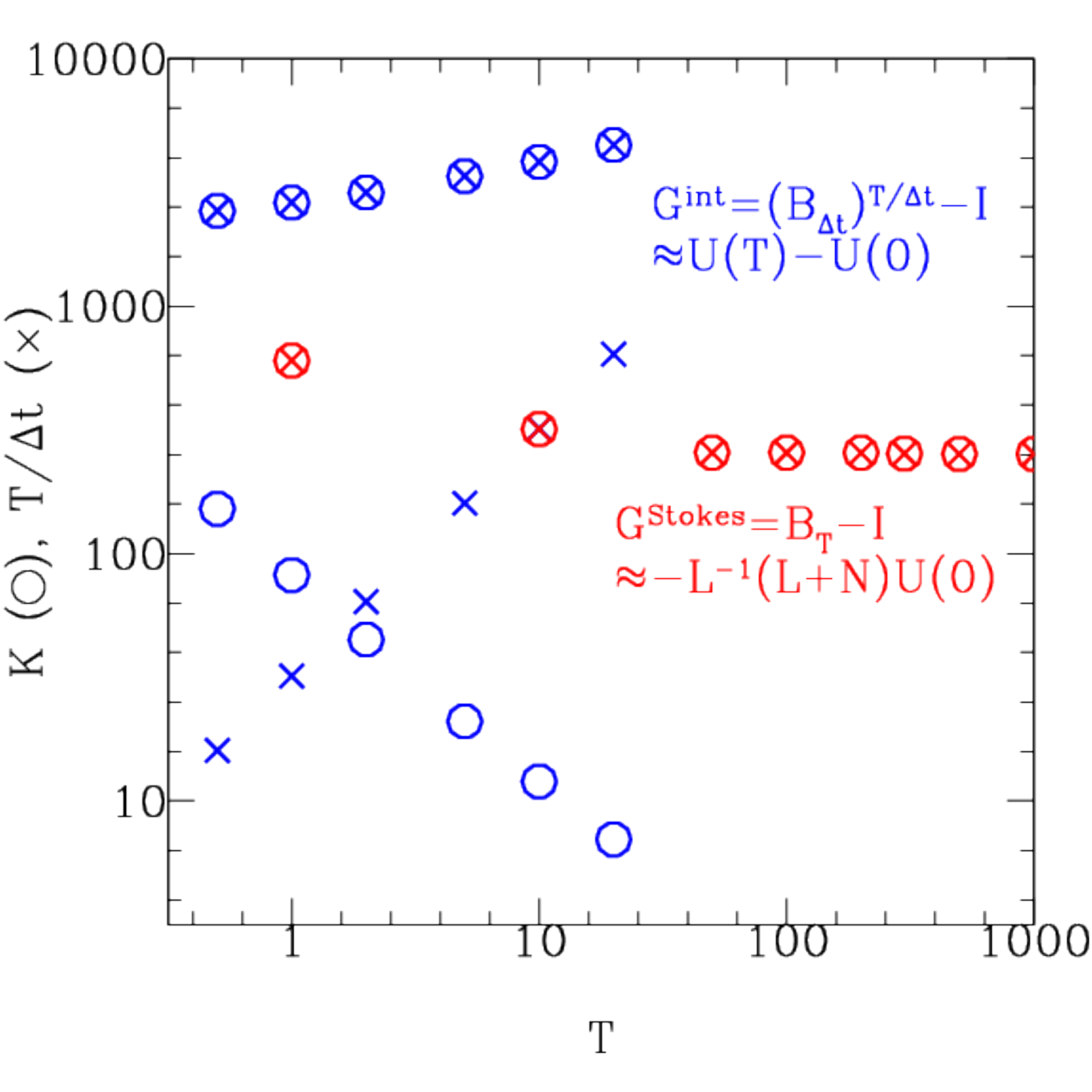}
\includegraphics[width=0.5\textwidth]{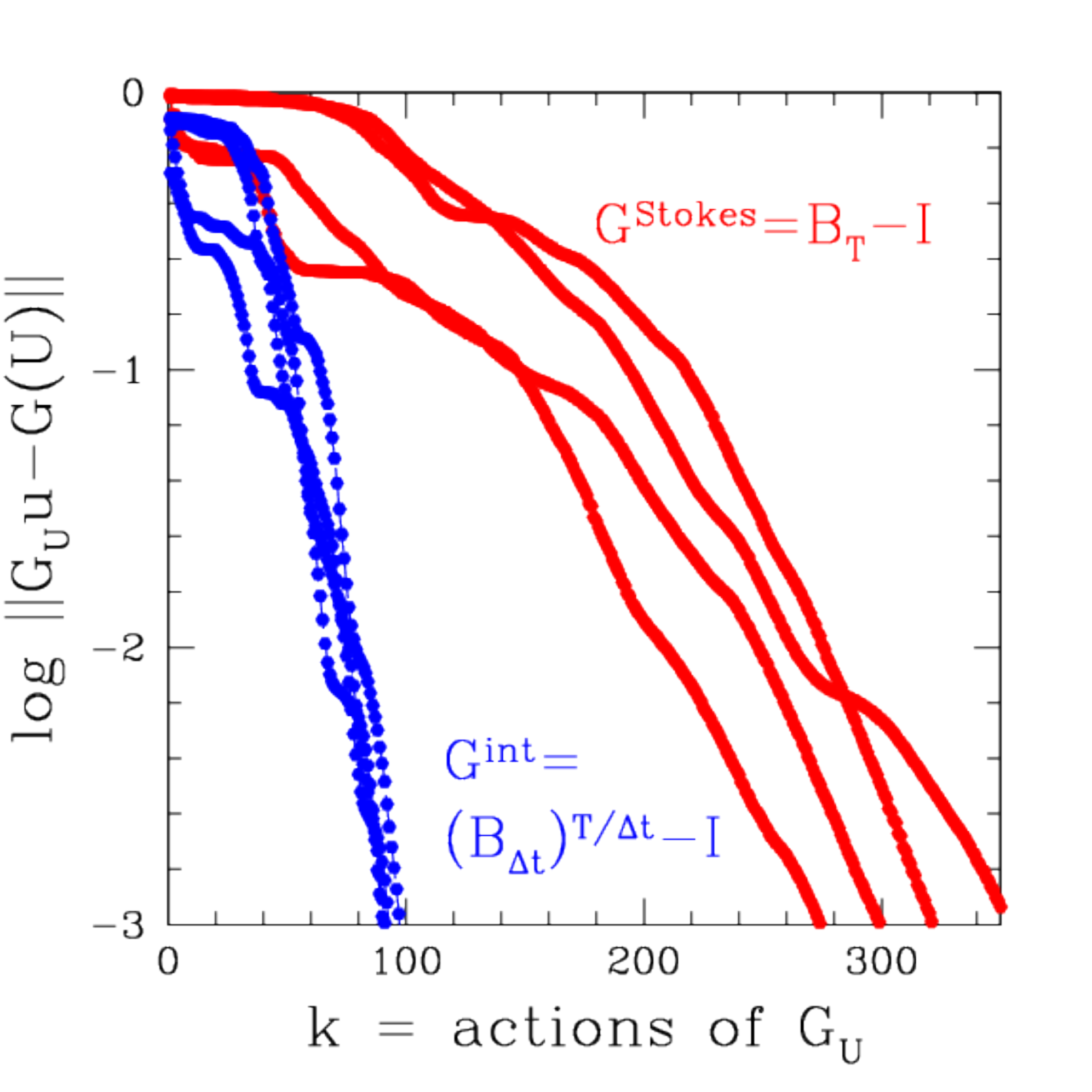}
\caption{	Left: Dependence of the number $K$ of GMRES iterations (hollow circles)
  required for convergence for a typical Newton step as a function of $T$,
  plotted along with the number $T/\dt$ of timesteps (crosses) per action of the operator.
  As $T$ increases, the number $T/\dt$ of timesteps required per
  action obviously increases, while the $K$ required by $\Gint$ (hollow circles)
  decreases from 152 to 7, showing that its condition number improves.
  In contrast, for $\Gstokes$, $T/\dt=1$ but the $K$ required 
  is much larger, decreasing with $T$ and saturating at around 230 for $T\gtrsim 10$.
  Circles containing crosses shows $\log(K)+\log(T/\dt)= \log(KT/\dt)$,
  the log of the total number of timesteps in a typical Newton step.
  Right: Convergence during the solution of linear equations
  for four successive Newton steps at $Re=500$ for $\Gint_{T=1}$
  and for $\Gstokes_{T=100}$. For each curve, convergence is
  interrupted by plateaus. The linear system
  involving $\Gint$ requires fewer actions with $G_U$
  than $\Gstokes$, a consequence of the fact that $\Gint$ is
  better conditioned than $\Gstokes$.
  The number $K$ of actions required by $\Gstokes$ increases
  at each successive Newton step.
}
\label{fig:GMRES_scaling_and_convergence}
\end{figure}


Many steady states of plane Couette flow have been computed using Channelflow;
see \cite{gibson2009equilibrium}.
Here, we investigate a branch called EQ1 or NCBW, since it was discovered
separately by Nagata \cite{Nagata1990three}, by Clever \& Busse
\cite{clever1997tertiary}
and by Waleffe \cite{waleffe1998three,waleffe2003homotopy}. 
It is illustrated in Fig.\ \ref{fig:EQ1NCBW}.
We begin from a member of EQ1 at $Re=400$ and lower the Reynolds number
to $Re=380$.  We use a domain of size $2\pi/1.14 \times 2 \times 2\pi / 2.5$
with a numerical resolution of $(M_x, M_y, Mz) = (48, 35, 48)$.
At each Newton iteration, the L$_2$ norm $||G(U)||$ is measured.
Each Newton step requires the solution of a linear system, which in turn
requires $K\times T/\dt$ timesteps. We plot $||G(U)||$ as a function of the
timesteps executed.  The left portion of Fig.~\ref{fig:classic_and_stokes} shows
that the best performance for the multi-step time-integration-based $\Gint$
with $\dt=0.031\,25$ is achieved for $T\lesssim 1$, while the right portion
shows that the best performance for the single-step
Stokes-preconditioned $\Gstokes$ is achieved for $\dt=T\gtrsim 100$.  The most
striking result of Fig.~\ref{fig:classic_and_stokes}, however, is that the Stokes
method takes about 10\% of the number of timesteps of the classic integration
method.

In order to better understand the performance of these two methods, 
we recall that the number of timesteps per Newton iteration is the product of $T/\dt$ and $K$.
For $\Gint$, $T/\dt$ obviously increases with the total integration time $T$,
while $K$ decreases with $T$, i.e. $\Gint$ becomes better conditioned.
Figure \ref{fig:GMRES_scaling_and_convergence} (left) illustrates this tradeoff.
The two effects combine to show a slight increase in number of timesteps
with final time $T$.
In contrast, for $\Gstokes$, the number $K$
of operator actions needed is far higher than is needed for $\Gint$, 
but only a single timestep is needed per operator action. 
This leads to a far lower cost for $\Gstokes$ than for $\Gint$.
For $\Gstokes$, increasing $T$ makes the preconditioning more effective,
leading to a decrease in $K$ and therefore in the number of timesteps taken
as shown in the left portions of Fig.~\ref{fig:classic_and_stokes} and
\ref{fig:GMRES_scaling_and_convergence}. The decrease is monotonic
and saturates at around $T=100$. (However, in \cite{beaume2017adaptive}
a local minimum is found, with optimal performance at $T=0.1$.)
We emphasize that changing $T$ in $\Gstokes$ affects only the rate of
convergence. In contrast, the roots themselves of $\Gint$ necessarily depend
somewhat on $\dt$ and, to a lesser extent, on $T$ as well. 
Figure \ref{fig:GMRES_scaling_and_convergence} (right) shows the
convergence of the solution of the linear systems arising in the
successive Newton steps at $Re=500$ for $\Gint_{T=1}$ and $\Gstokes_{T=100}$.
Convergence is interrupted by
plateaus, whose understanding  might lead to better preconditioners.
Note that $K$ increases from one Newton step to the next, which is especially
noticeable for $\Gstokes$.

We have calculated the EQ1 branch from $Re=400$ up to $Re=3000$.
The conditioning of both operators worsens:
for $Re=2000$, the number of timesteps required
by the Stokes preconditioning and the integration methods 
is about 4000 and 40\,000, respectively, 
while for $Re=3000$, it is about 9000 and 90\,000.
Thus, the time required increases for both methods, but the
ratio between the two methods in terms of number of timesteps
remains approximately the same: on the order of 10.
For the case of a lid-driven cavity, Brynjell-Rahkola et al.
\cite{brynjell2017computing} show that the number of GMRES
iterations required by $\Gstokes$ using Nek5000 \cite{Nekton}
is approximately
proportional to $Re$, as $Re$ is increased from 100 to 500.
It was previously shown \cite{tuckerman2015laplacian}
that the condition number of $\Gstokes$ is unaffected
by an increase in spatial resolution.

\subsection{Computations with Openpipeflow}

\begin{figure}
\includegraphics[width=0.45\textwidth]{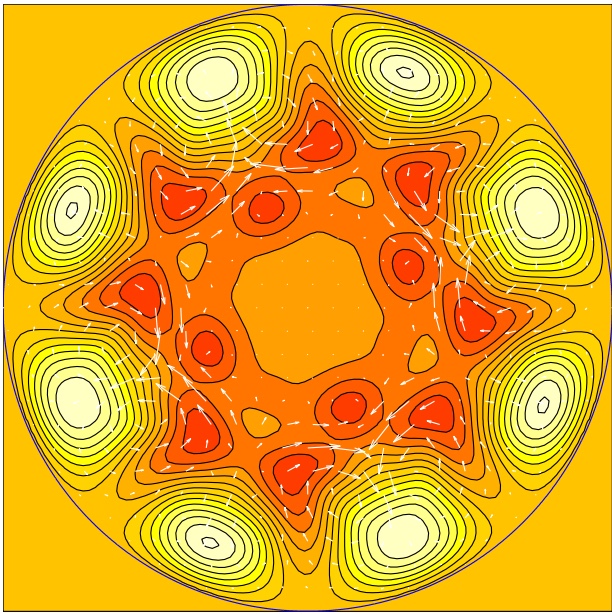}\qquad
\includegraphics[width=0.45\textwidth]{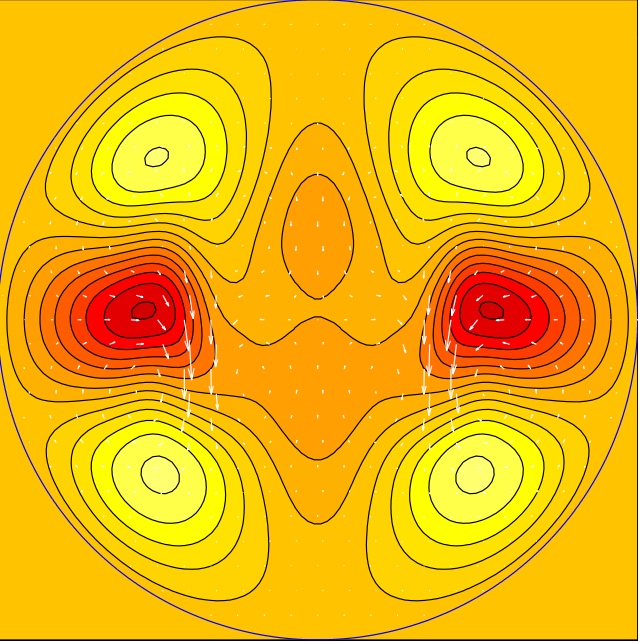}
\caption{Traveling wave states N4L (left) and M1L (right),
  visualized via the deviation of the streamwise velocity from the laminar profile.
From Pringle, Duguet and Kerswell \cite{pringle2009highly}.}
\label{fig:pipes}
\end{figure}

%
\begin{figure}
\includegraphics[width=0.5\textwidth]{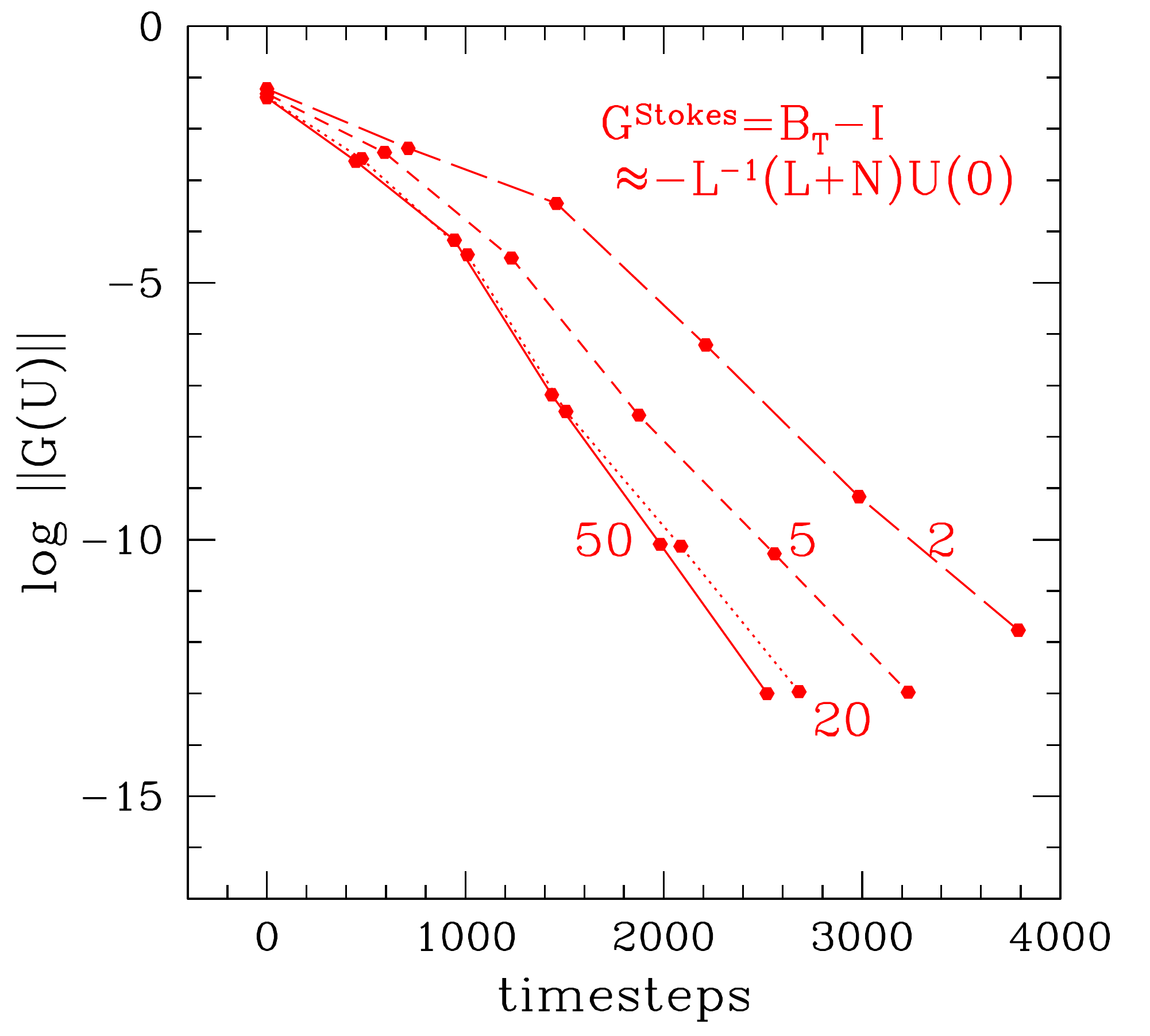}
\includegraphics[width=0.5\textwidth]{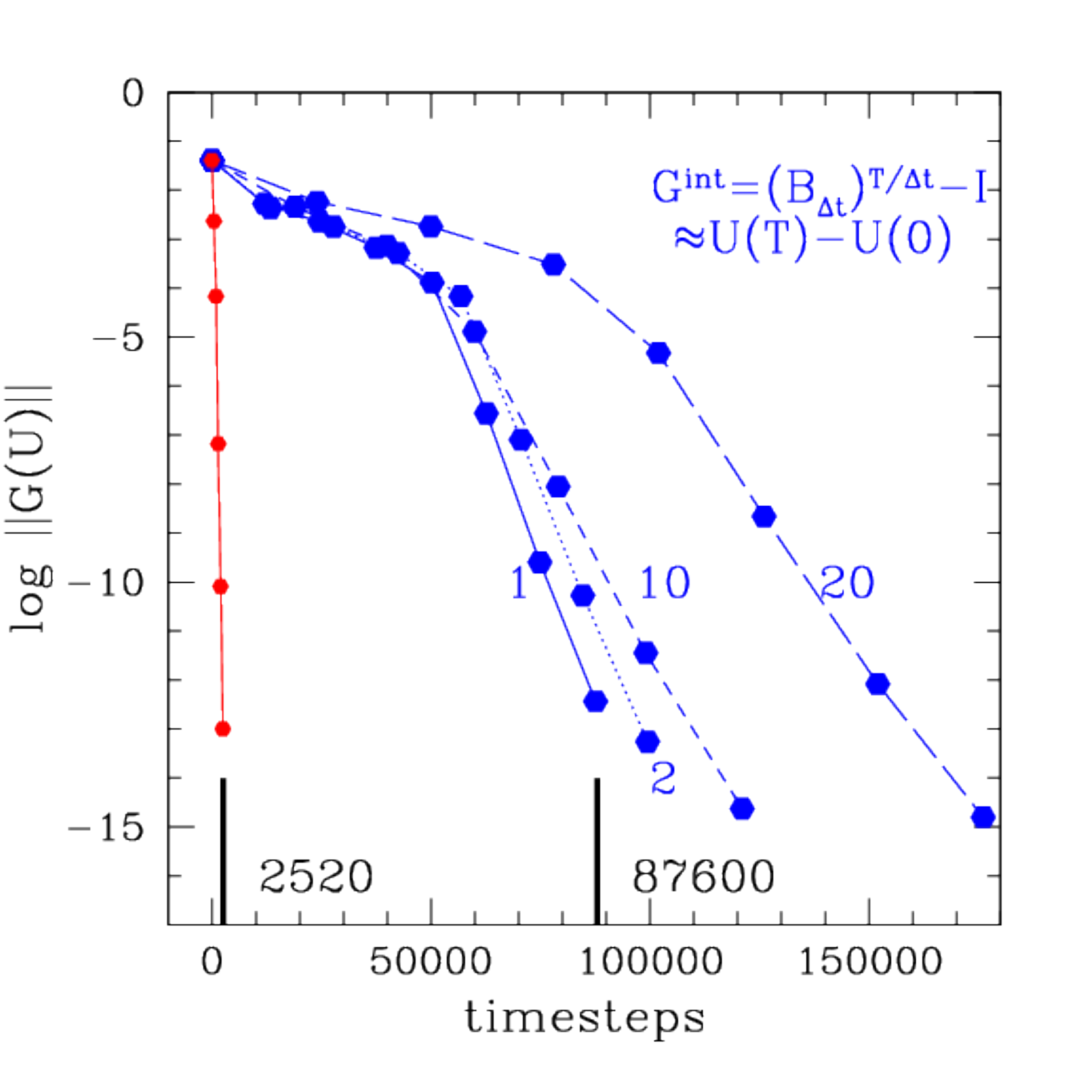}
\caption{Performance of Newton's method with GMRES in Openpipeflow for 
  single-step Stokes preconditioning method $\Gstokes_T$ and for
  multi-step integration method $\Gint_T$ for various values of $T$.
  The initial condition is the N4L traveling wave state 
  at $Re=2500$ and $Re$ is raised to $Re=2600$.
  Ordinate shows $\log||G(U)||$ at each Newton iteration, while abscissa
  shows the number of timesteps taken thus far.
  Each curve is labelled by its value of $T$.
 Left: Performance of single-step $\Gstokes_T \equiv B_T-I$.
  Convergence is fastest when $T$ is highest, achieving
  approximately asymptotic performance for $T=50$ (solid curve).
  Right: 
Performance of multi-step $\Gint_T \equiv (B_{\dt})^{T/\dt}-I$
  with timestep $\dt=0.01$.
  Convergence is fastest when $T$ is lowest, achieving approximately
  asymptotic performance for $T=1$ (solid curve). Shown for comparison is the
  convergence curve for $\Gstokes$ with $T=50$.
  Short black lines highlight the difference in speed between the two methods:
  the classic integration method takes $87\,600$ timesteps, while 
  the Stokes method takes $2520$ timesteps, a ratio of about $35$.}
\label{fig:N4L_2600}
\end{figure}
\begin{figure}
\includegraphics[width=0.5\textwidth]{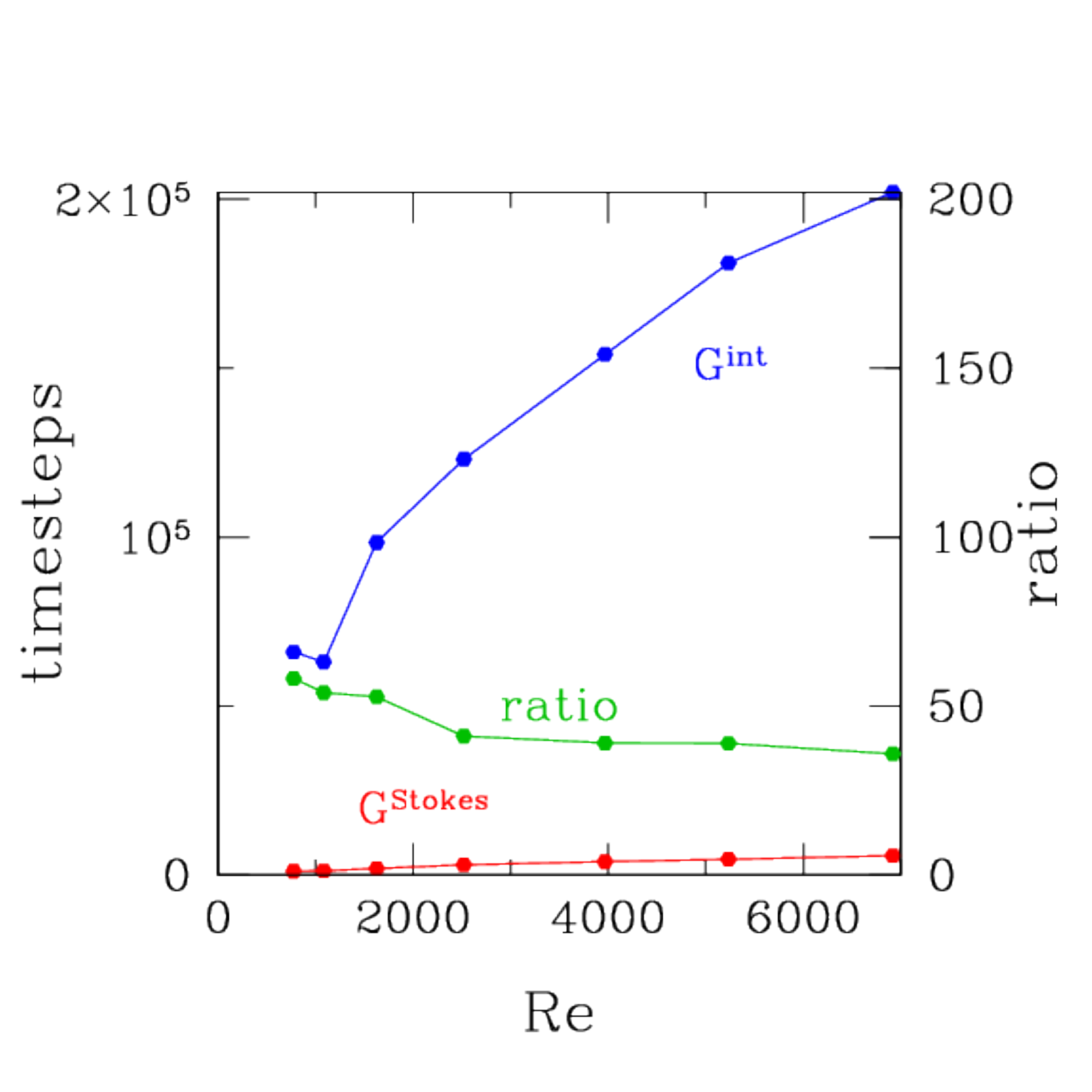}
\includegraphics[width=0.5\textwidth]{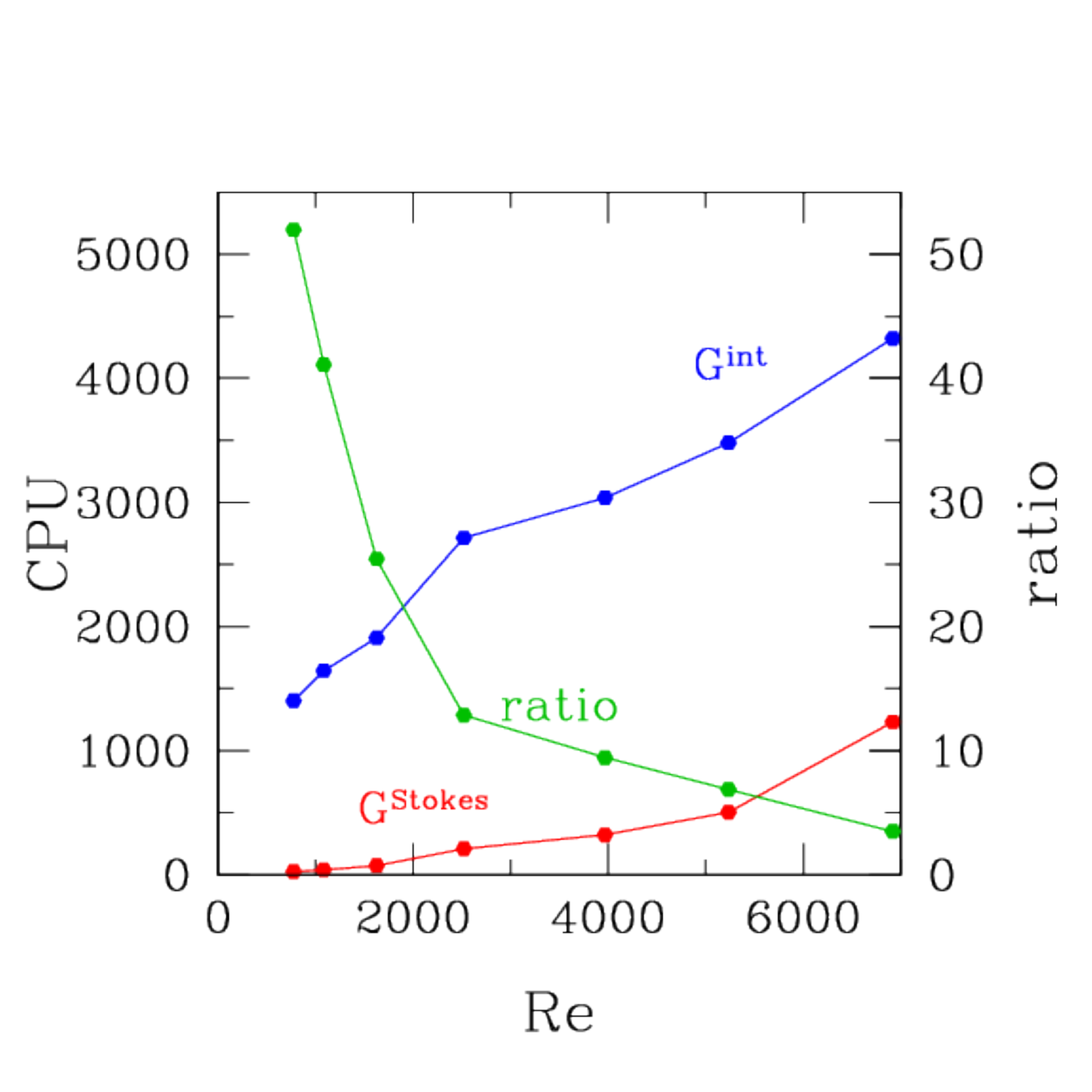}
\caption{Time taken to compute the highly symmetric M1L state as a
  function of Reynolds number. Comparison in terms of number of
  timesteps (left) and CPU time in seconds (right).
  Computation with single-step $\Gstokes_T$ uses $T=50$
  while computation with multi-step $\Gint_T$ uses $T=5$.
  The time taken to calculate the traveling waves 
  increases for both the Stokes and the integration method. The ratio
  of timesteps between the two methods, about 50, changes very little,
  while the ratio of CPU times decreases from 55 to about 2. }
\label{fig:M1L}
\end{figure}

We now turn to calculations carried out using Openpipeflow. 
Since there is a mean flow along the streamwise direction $x$ of the pipe,
bifurcations breaking the $x$ symmetry are always to traveling waves. 
Visualizations of two such states, namely the N4L and M1L waves computed
by Pringle, Duguet and Kerswell \cite{pringle2009highly}, are given in
Fig.~\ref{fig:pipes}. Figure \ref{fig:N4L_2600} presents 
the convergence of the calculation of N4L at $Re=2600$ with Openpipeflow,
starting from the solution at $Re=2500$. 
(We recall that Reynolds numbers for pipe flow are defined in such a way that that
they are 4 to 6 times the equivalent Reynolds numbers for plane Couette flow.)
The axial length is $\pi/1.7$ diameters and the resolution is
$(M_r,M_\theta,M_z)=(64,36,54)$.
 Measured in terms of the number of timesteps, the calculation using
$\Gstokes$ is 35 times faster than that using $\Gint$.
We then calculate states along the highly symmetric M1L branch as a function of
Reynolds number.
The axial length is $\pi/1.437$ diameters and the resolution is
$(M_r,M_\theta,M_z)=(64,48,54)$.
The speedup is even more dramatic, reaching a ratio of 50
in terms of timesteps between the Stokes and integration methods.
This ratio remains approximately constant as the Reynolds number is increased,
as shown in the left portion of Fig.~\ref{fig:M1L}.

In the right portion of Fig.~\ref{fig:M1L} we contrast 
the CPU time in seconds required by the two methods on a single processor of a Xeon
X5650@2.67GHz. The Stokes method remains
much faster than the integration method, but far less 
than when it is measured in terms of number of timesteps.
The reason for this has been alluded to before, in Eq.~\eqref{eq:CPU}.
For sufficiently large $K$ (which occurs at sufficiently high $Re$),
the CPU time taken by GMRES to orthogonalize the Krylov vectors
to one another, which scales like $K^2$, becomes comparable to and even exceeds
the CPU time taken by the operator actions.

\subsection{Operator Spectra}

\begin{figure}
\begingroup
\newcommand{\keyfont}[0]{\small}
  \makeatletter
  \providecommand\color[2][]{%
    \GenericError{(gnuplot) \space\space\space\@spaces}{%
      Package color not loaded in conjunction with
      terminal option `colourtext'%
    }{See the gnuplot documentation for explanation.%
    }{Either use 'blacktext' in gnuplot or load the package
      color.sty in LaTeX.}%
    \renewcommand\color[2][]{}%
  }%
  \providecommand\includegraphics[2][]{%
    \GenericError{(gnuplot) \space\space\space\@spaces}{%
      Package graphicx or graphics not loaded%
    }{See the gnuplot documentation for explanation.%
    }{The gnuplot epslatex terminal needs graphicx.sty or graphics.sty.}%
    \renewcommand\includegraphics[2][]{}%
  }%
  \providecommand\rotatebox[2]{#2}%
  \@ifundefined{ifGPcolor}{%
    \newif\ifGPcolor
    \GPcolortrue
  }{}%
  \@ifundefined{ifGPblacktext}{%
    \newif\ifGPblacktext
    \GPblacktexttrue
  }{}%
  \let\gplgaddtomacro\g@addto@macro
  \gdef\gplbacktext{}%
  \gdef\gplfronttext{}%
  \makeatother
  \ifGPblacktext
    \def\colorrgb#1{}%
    \def\colorgray#1{}%
  \else
    \ifGPcolor
      \def\colorrgb#1{\color[rgb]{#1}}%
      \def\colorgray#1{\color[gray]{#1}}%
      \expandafter\def\csname LTw\endcsname{\color{white}}%
      \expandafter\def\csname LTb\endcsname{\color{black}}%
      \expandafter\def\csname LTa\endcsname{\color{black}}%
      \expandafter\def\csname LT0\endcsname{\color[rgb]{1,0,0}}%
      \expandafter\def\csname LT1\endcsname{\color[rgb]{0,1,0}}%
      \expandafter\def\csname LT2\endcsname{\color[rgb]{0,0,1}}%
      \expandafter\def\csname LT3\endcsname{\color[rgb]{1,0,1}}%
      \expandafter\def\csname LT4\endcsname{\color[rgb]{0,1,1}}%
      \expandafter\def\csname LT5\endcsname{\color[rgb]{1,1,0}}%
      \expandafter\def\csname LT6\endcsname{\color[rgb]{0,0,0}}%
      \expandafter\def\csname LT7\endcsname{\color[rgb]{1,0.3,0}}%
      \expandafter\def\csname LT8\endcsname{\color[rgb]{0.5,0.5,0.5}}%
    \else
      \def\colorrgb#1{\color{black}}%
      \def\colorgray#1{\color[gray]{#1}}%
      \expandafter\def\csname LTw\endcsname{\color{white}}%
      \expandafter\def\csname LTb\endcsname{\color{black}}%
      \expandafter\def\csname LTa\endcsname{\color{black}}%
      \expandafter\def\csname LT0\endcsname{\color{black}}%
      \expandafter\def\csname LT1\endcsname{\color{black}}%
      \expandafter\def\csname LT2\endcsname{\color{black}}%
      \expandafter\def\csname LT3\endcsname{\color{black}}%
      \expandafter\def\csname LT4\endcsname{\color{black}}%
      \expandafter\def\csname LT5\endcsname{\color{black}}%
      \expandafter\def\csname LT6\endcsname{\color{black}}%
      \expandafter\def\csname LT7\endcsname{\color{black}}%
      \expandafter\def\csname LT8\endcsname{\color{black}}%
    \fi
  \fi
  \setlength{\unitlength}{0.0500bp}%
  \begin{picture}(6660.00,6660.00)%
    \gplgaddtomacro\gplbacktext{%
      \csname LTb\endcsname%
      \put(408,1013){\makebox(0,0)[r]{\strut{}\keyfont $-1$}}%
      \csname LTb\endcsname%
      \put(408,1851){\makebox(0,0)[r]{\strut{}\keyfont $0$}}%
      \csname LTb\endcsname%
      \put(408,2689){\makebox(0,0)[r]{\strut{}\keyfont $1$}}%
      \csname LTb\endcsname%
      \put(510,408){\makebox(0,0){\strut{}\keyfont $-2$}}%
      \csname LTb\endcsname%
      \put(1348,408){\makebox(0,0){\strut{}\keyfont $-1$}}%
      \csname LTb\endcsname%
      \put(2185,408){\makebox(0,0){\strut{}\keyfont $0$}}%
      \csname LTb\endcsname%
      \put(3023,408){\makebox(0,0){\strut{}\keyfont $1$}}%
    }%
    \gplgaddtomacro\gplfronttext{%
    }%
    \gplgaddtomacro\gplbacktext{%
      \csname LTb\endcsname%
      \put(4088,419){\makebox(0,0){\strut{}\keyfont $-1.001$}}%
      \csname LTb\endcsname%
      \put(4780,419){\makebox(0,0){\strut{}\keyfont $-1$}}%
      \csname LTb\endcsname%
      \put(5472,419){\makebox(0,0){\strut{}\keyfont $-0.999$}}%
      \csname LTb\endcsname%
      \put(6128,1159){\makebox(0,0)[l]{\strut{}$-10^{-3}$}}%
      \csname LTb\endcsname%
      \put(6128,1851){\makebox(0,0)[l]{\strut{}$0$}}%
      \csname LTb\endcsname%
      \put(6128,2543){\makebox(0,0)[l]{\strut{}$10^{-3}$}}%
    }%
    \gplgaddtomacro\gplfronttext{%
    }%
    \gplgaddtomacro\gplbacktext{%
      \csname LTb\endcsname%
      \put(408,4828){\makebox(0,0)[r]{\strut{}\keyfont $-25$}}%
      \csname LTb\endcsname%
      \put(408,5181){\makebox(0,0)[r]{\strut{}\keyfont $0$}}%
      \csname LTb\endcsname%
      \put(408,5533){\makebox(0,0)[r]{\strut{}\keyfont $25$}}%
      \csname LTb\endcsname%
      \put(862,4642){\makebox(0,0){\strut{}\keyfont $-350$}}%
      \csname LTb\endcsname%
      \put(1567,4642){\makebox(0,0){\strut{}\keyfont $-300$}}%
      \csname LTb\endcsname%
      \put(2272,4642){\makebox(0,0){\strut{}\keyfont $-250$}}%
      \csname LTb\endcsname%
      \put(2977,4642){\makebox(0,0){\strut{}\keyfont $-200$}}%
      \csname LTb\endcsname%
      \put(67,5993){\makebox(0,0)[l]{\strut{}(a)}}%
      \put(67,3130){\makebox(0,0)[l]{\strut{}(b)}}%
      \put(3196,3130){\makebox(0,0)[l]{\strut{}(c)}}%
    }%
    \gplgaddtomacro\gplfronttext{%
      \csname LTb\endcsname%
      \put(2281,6334){\makebox(0,0)[r]{\strut{}\keyfont $G^{\text{Stokes}}_{0.0001}$}}%
      \csname LTb\endcsname%
      \put(2281,5962){\makebox(0,0)[r]{\strut{}\keyfont $G^{\text{Stokes}}_{100}$}}%
      \csname LTb\endcsname%
      \put(3375,6334){\makebox(0,0)[r]{\strut{}\keyfont $G^{\text{Stokes}}_{0.1}$}}%
      \csname LTb\endcsname%
      \put(3375,5962){\makebox(0,0)[r]{\strut{}\keyfont$G^{\text{int}}_{10}$}}%
    }%
    \gplgaddtomacro\gplbacktext{%
      \csname LTb\endcsname%
      \put(4311,3516){\makebox(0,0){\strut{}\keyfont $-40$}}%
      \csname LTb\endcsname%
      \put(5051,3516){\makebox(0,0){\strut{}\keyfont $-20$}}%
      \csname LTb\endcsname%
      \put(5790,3516){\makebox(0,0){\strut{}\keyfont $0$}}%
      \csname LTb\endcsname%
      \put(6262,3702){\makebox(0,0)[l]{\strut{}\keyfont $-40$}}%
      \csname LTb\endcsname%
      \put(6262,4441){\makebox(0,0)[l]{\strut{}\keyfont $-20$}}%
      \csname LTb\endcsname%
      \put(6262,5181){\makebox(0,0)[l]{\strut{}\keyfont $0$}}%
      \csname LTb\endcsname%
      \put(6262,5920){\makebox(0,0)[l]{\strut{}\keyfont $20$}}%
      \csname LTb\endcsname%
      \put(6262,6659){\makebox(0,0)[l]{\strut{}\keyfont $40$}}%
      \csname LTb\endcsname%
      \put(67,5993){\makebox(0,0)[l]{\strut{}(a)}}%
      \put(67,3130){\makebox(0,0)[l]{\strut{}(b)}}%
      \put(3196,3130){\makebox(0,0)[l]{\strut{}(c)}}%
      \put(3682,5217){\makebox(0,0)[l]{\strut{}$\ldots$}}%
    }%
    \gplgaddtomacro\gplfronttext{%
    }%
    \gplbacktext
    \put(0,0){\includegraphics{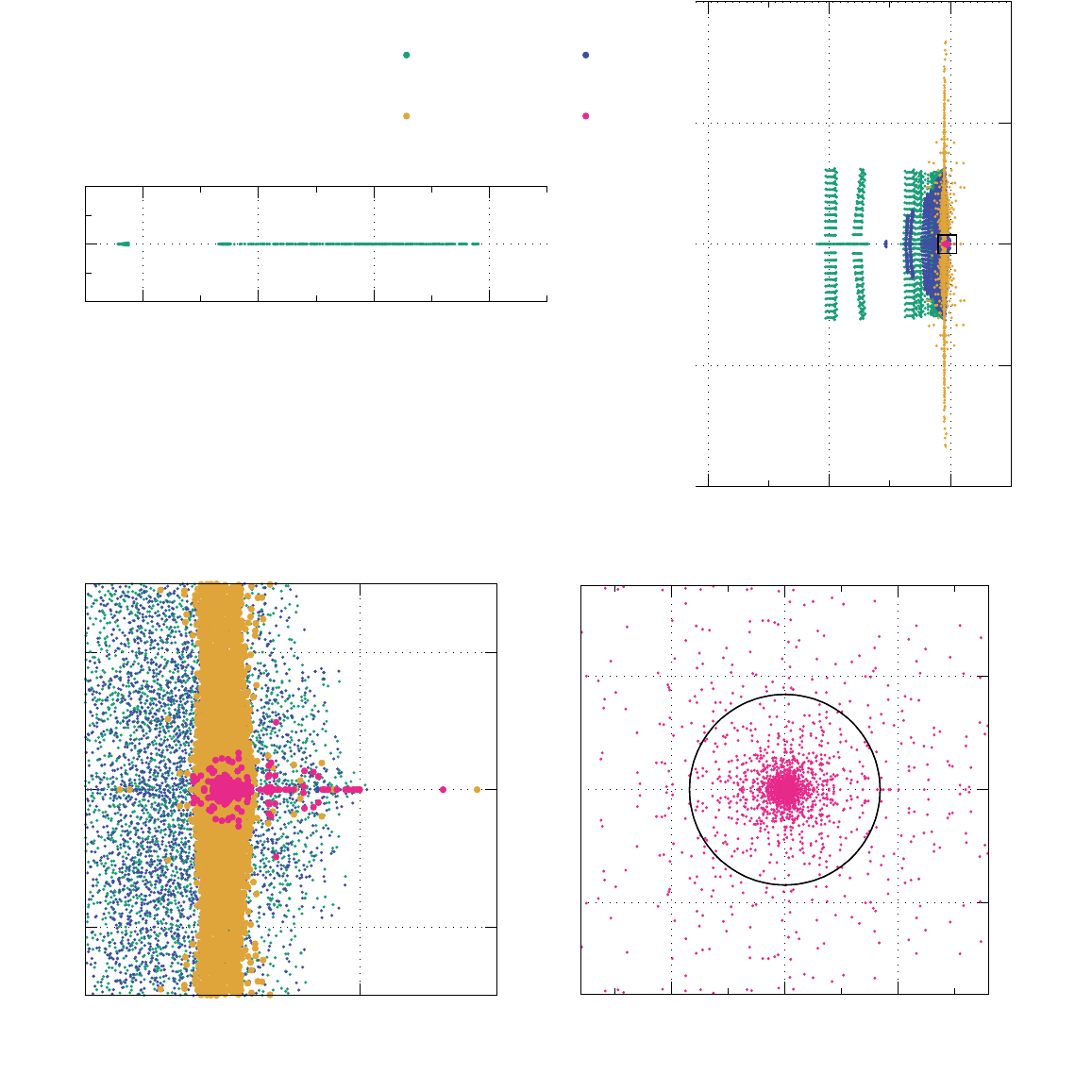}}%
    \gplfronttext
  \end{picture}%
\endgroup
\caption{Spectra in the complex plane of the single-step operators $\Gstokes_{0.0001}$ (green),
  $\Gstokes_{0.1}$ (blue), $\Gstokes_{100}$ (ochre) and the multi-step operator
  $\Gint_{10}$ (magenta) for EQ1 at $Re=500$.
(a)  The very large negative real eigenvalues of $\Gstokes_{0.001}$
  near $-360$ are not present in $\Gstokes_{0.1}$, 
  for which the dominant eigenvalues have real parts near $-10$ and
  imaginary parts near $\pm 12$. When $T$ is further increased to 100,
  most of the real parts cluster within $\pm 0.2$ of $-1$, while the
  imaginary parts extend to $\pm 35$.
  (b) Close-up showing the spectra near the point $(-1,0)$. The plotted area is
  indicated by an overlaid black square in part (a). At this scale, the
  spectra of the single-step operators are 
  so close as to be indistinguishable, save for a few
  outliers. Consequently, $\Gstokes_{100}$ and $\Gint_{10}$ have been plotted
  with larger points for emphasis and only every third point from
  $\Gstokes_{0.0001}$ and $\Gstokes_{0.1}$ has been included. 
  The outlying $\Gint_{10}$ point near $0.6$ is approximately $T=10$ times the
      leading eigenvalue \cite{wang2007lower} of $L+N_U$. 
(c) Highly enlarged plot showing the spectrum of $\Gint_{10}$ only.
  Most eigenvalues of this well-conditioned multi-step
  operator cluster tightly around $-1$, since the black
	  circle of radius $8.4\times 10^{-4}$, centered at $(-1, 0)$, contains 95\%
	  of the spectrum.}
\label{fig:spectra500}
\end{figure}
\begin{figure}
\begingroup
\newcommand{\keyfont}[0]{\small}
  \makeatletter
  \providecommand\color[2][]{%
    \GenericError{(gnuplot) \space\space\space\@spaces}{%
      Package color not loaded in conjunction with
      terminal option `colourtext'%
    }{See the gnuplot documentation for explanation.%
    }{Either use 'blacktext' in gnuplot or load the package
      color.sty in LaTeX.}%
    \renewcommand\color[2][]{}%
  }%
  \providecommand\includegraphics[2][]{%
    \GenericError{(gnuplot) \space\space\space\@spaces}{%
      Package graphicx or graphics not loaded%
    }{See the gnuplot documentation for explanation.%
    }{The gnuplot epslatex terminal needs graphicx.sty or graphics.sty.}%
    \renewcommand\includegraphics[2][]{}%
  }%
  \providecommand\rotatebox[2]{#2}%
  \@ifundefined{ifGPcolor}{%
    \newif\ifGPcolor
    \GPcolortrue
  }{}%
  \@ifundefined{ifGPblacktext}{%
    \newif\ifGPblacktext
    \GPblacktexttrue
  }{}%
  \let\gplgaddtomacro\g@addto@macro
  \gdef\gplbacktext{}%
  \gdef\gplfronttext{}%
  \makeatother
  \ifGPblacktext
    \def\colorrgb#1{}%
    \def\colorgray#1{}%
  \else
    \ifGPcolor
      \def\colorrgb#1{\color[rgb]{#1}}%
      \def\colorgray#1{\color[gray]{#1}}%
      \expandafter\def\csname LTw\endcsname{\color{white}}%
      \expandafter\def\csname LTb\endcsname{\color{black}}%
      \expandafter\def\csname LTa\endcsname{\color{black}}%
      \expandafter\def\csname LT0\endcsname{\color[rgb]{1,0,0}}%
      \expandafter\def\csname LT1\endcsname{\color[rgb]{0,1,0}}%
      \expandafter\def\csname LT2\endcsname{\color[rgb]{0,0,1}}%
      \expandafter\def\csname LT3\endcsname{\color[rgb]{1,0,1}}%
      \expandafter\def\csname LT4\endcsname{\color[rgb]{0,1,1}}%
      \expandafter\def\csname LT5\endcsname{\color[rgb]{1,1,0}}%
      \expandafter\def\csname LT6\endcsname{\color[rgb]{0,0,0}}%
      \expandafter\def\csname LT7\endcsname{\color[rgb]{1,0.3,0}}%
      \expandafter\def\csname LT8\endcsname{\color[rgb]{0.5,0.5,0.5}}%
    \else
      \def\colorrgb#1{\color{black}}%
      \def\colorgray#1{\color[gray]{#1}}%
      \expandafter\def\csname LTw\endcsname{\color{white}}%
      \expandafter\def\csname LTb\endcsname{\color{black}}%
      \expandafter\def\csname LTa\endcsname{\color{black}}%
      \expandafter\def\csname LT0\endcsname{\color{black}}%
      \expandafter\def\csname LT1\endcsname{\color{black}}%
      \expandafter\def\csname LT2\endcsname{\color{black}}%
      \expandafter\def\csname LT3\endcsname{\color{black}}%
      \expandafter\def\csname LT4\endcsname{\color{black}}%
      \expandafter\def\csname LT5\endcsname{\color{black}}%
      \expandafter\def\csname LT6\endcsname{\color{black}}%
      \expandafter\def\csname LT7\endcsname{\color{black}}%
      \expandafter\def\csname LT8\endcsname{\color{black}}%
    \fi
  \fi
  \setlength{\unitlength}{0.0500bp}%
  \begin{picture}(6660.00,6660.00)%
    \gplgaddtomacro\gplbacktext{%
      \csname LTb\endcsname%
      \put(408,1013){\makebox(0,0)[r]{\strut{}\keyfont $-1$}}%
      \csname LTb\endcsname%
      \put(408,1851){\makebox(0,0)[r]{\strut{}\keyfont $0$}}%
      \csname LTb\endcsname%
      \put(408,2689){\makebox(0,0)[r]{\strut{}\keyfont $1$}}%
      \csname LTb\endcsname%
      \put(510,408){\makebox(0,0){\strut{}\keyfont $-2$}}%
      \csname LTb\endcsname%
      \put(1348,408){\makebox(0,0){\strut{}\keyfont $-1$}}%
      \csname LTb\endcsname%
      \put(2185,408){\makebox(0,0){\strut{}\keyfont $0$}}%
      \csname LTb\endcsname%
      \put(3023,408){\makebox(0,0){\strut{}\keyfont $1$}}%
    }%
    \gplgaddtomacro\gplfronttext{%
    }%
    \gplgaddtomacro\gplbacktext{%
      \csname LTb\endcsname%
      \put(3949,419){\makebox(0,0){\strut{}\keyfont $-1.2$}}%
      \csname LTb\endcsname%
      \put(4780,419){\makebox(0,0){\strut{}\keyfont $-1$}}%
      \csname LTb\endcsname%
      \put(5611,419){\makebox(0,0){\strut{}\keyfont $-0.8$}}%
      \csname LTb\endcsname%
      \put(6128,1020){\makebox(0,0)[l]{\strut{}\keyfont $-0.2$}}%
      \csname LTb\endcsname%
      \put(6128,1851){\makebox(0,0)[l]{\strut{}\keyfont $0$}}%
      \csname LTb\endcsname%
      \put(6128,2682){\makebox(0,0)[l]{\strut{}\keyfont $0.2$}}%
    }%
    \gplgaddtomacro\gplfronttext{%
    }%
    \gplgaddtomacro\gplbacktext{%
      \csname LTb\endcsname%
      \put(1861,3702){\makebox(0,0)[r]{\strut{}\keyfont $-25$}}%
      \csname LTb\endcsname%
      \put(1861,4293){\makebox(0,0)[r]{\strut{}\keyfont $0$}}%
      \csname LTb\endcsname%
      \put(1861,4885){\makebox(0,0)[r]{\strut{}\keyfont $25$}}%
      \csname LTb\endcsname%
      \put(1861,5476){\makebox(0,0)[r]{\strut{}\keyfont $50$}}%
      \csname LTb\endcsname%
      \put(1861,6068){\makebox(0,0)[r]{\strut{}\keyfont $75$}}%
      \csname LTb\endcsname%
      \put(1861,6659){\makebox(0,0)[r]{\strut{}\keyfont $100$}}%
      \csname LTb\endcsname%
      \put(1963,3516){\makebox(0,0){\strut{}\keyfont $-125$}}%
      \csname LTb\endcsname%
      \put(2555,3516){\makebox(0,0){\strut{}\keyfont $-100$}}%
      \csname LTb\endcsname%
      \put(3146,3516){\makebox(0,0){\strut{}\keyfont $-75$}}%
      \csname LTb\endcsname%
      \put(3738,3516){\makebox(0,0){\strut{}\keyfont $-50$}}%
      \csname LTb\endcsname%
      \put(4329,3516){\makebox(0,0){\strut{}\keyfont $-25$}}%
      \csname LTb\endcsname%
      \put(4921,3516){\makebox(0,0){\strut{}\keyfont $0$}}%
      \csname LTb\endcsname%
      \put(5512,3516){\makebox(0,0){\strut{}\keyfont $25$}}%
      \csname LTb\endcsname%
      \put(1065,6659){\makebox(0,0)[l]{\strut{}(a)}}%
      \put(67,3130){\makebox(0,0)[l]{\strut{}(b)}}%
      \put(3196,3130){\makebox(0,0)[l]{\strut{}(c)}}%
    }%
    \gplgaddtomacro\gplfronttext{%
      \csname LTb\endcsname%
      \put(980,4808){\makebox(0,0)[r]{\strut{}\keyfont $G^{\text{Stokes}}_{0.0001}$}}%
      \csname LTb\endcsname%
      \put(980,4436){\makebox(0,0)[r]{\strut{}\keyfont $G^{\text{Stokes}}_{0.1}$}}%
      \csname LTb\endcsname%
      \put(980,4064){\makebox(0,0)[r]{\strut{}\keyfont $G^{\text{Stokes}}_{100}$}}%
      \csname LTb\endcsname%
      \put(980,3692){\makebox(0,0)[r]{\strut{}\keyfont$G^{\text{int}}_{10}$}}%
    }%
    \gplgaddtomacro\gplbacktext{%
      \colorrgb{1.00,1.00,1.00}%
      \put(1065,6659){\makebox(0,0)[l]{\strut{}(a)}}%
      \put(67,3130){\makebox(0,0)[l]{\strut{}(b)}}%
      \put(3196,3130){\makebox(0,0)[l]{\strut{}(c)}}%
    }%
    \gplgaddtomacro\gplfronttext{%
    }%
    \gplbacktext
    \put(0,0){\includegraphics{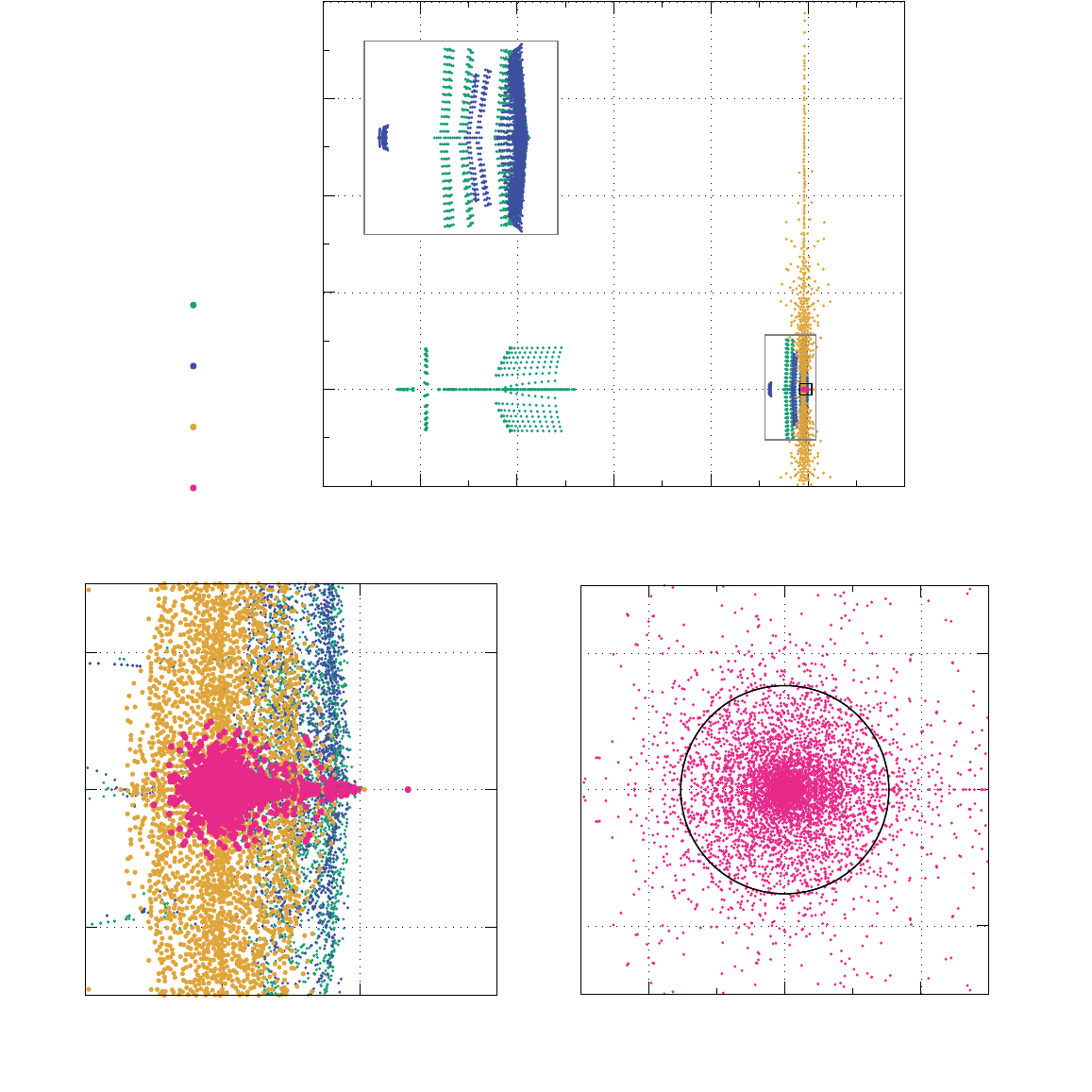}}%
    \gplfronttext
  \end{picture}%
\endgroup
\caption{Spectra in the complex plane of the single-step operators $\Gstokes_{0.0001}$ (green),
  $\Gstokes_{0.1}$ (blue), $\Gstokes_{100}$ (ochre) and the multi-step operator
  $\Gint_{10}$ (magenta) for EQ1 at $Re=1750$.
  (a) The most negative eigenvalue of $\Gstokes_{0.0001}$ is now near $-105$,
  while that of $\Gstokes_{0.1}$ is near $-10$.
  An enlargement showing the shape of these two spectra near the origin is
  included in the top left; its drawing region is indicated by the gray rectangle
  in the larger plot and its aspect ratio is $1:2$.
  The real parts of the eigenvalues of $\Gstokes_{100}$ continue
  to cluster around $-1$, but the imaginary parts now extend higher than in the
  $Re=500$ case, to $\pm 100$.
  (b) Close-up showing the spectra near the point $(-1,0)$. The plotted area is
  indicated by an overlaid black square in part (a). The spectra of
  $\Gint_{10}$ and $\Gstokes_{100}$ have been plotted with larger points to
  distinguish them from the  very dense regions of the $\Gstokes_{0.0001}$
  and $\Gstokes_{0.1}$ spectra, for which we include only every fourth point at
  this level of enlargement. 
  The outlying $\Gint_{10}$ point near $0.35$ is $T=10$ times the leading eigenvalue
  \cite{wang2007lower} of $L+N_U$.
  (c) Slightly closer enlargement showing only the $\Gint_{10}$ eigenvalues. The black
  circle of radius $0.153$, centered at $(-1, 0)$, contains $95\%$ of the
  spectrum. This should be contrasted with the $Re = 500$ case in which the
  eigenvalues are more tightly clustered by roughly two orders of magnitude.}
\label{fig:spectra1750}
\end{figure}

We have seen that the Stokes preconditioning method is much faster than the
integration-based method because the time required to act with $\Gstokes$ is
so much less than that required by $\Gint$. However, the conditioning of
$\Gstokes$ is much worse than that of $\Gint$. We want to understand why
this is so and also why the conditioning of both operators worsens as
the Reynolds number increases and improves as $T$ increases.
To do so, we calculate spectra of branch EQ1 of plane Couette flow using Channelflow.
This is done by constructing the full Jacobian matrix by acting with
$\Gint$ or $\Gstokes$ on successive basis vectors to form each column.
The eigenvalues are determined using Julia, which in
turn uses Lapack's geevx routine.

Figure \ref{fig:spectra500} shows the spectra at $Re=500$
of the Jacobians $G_U$ of operators $\Gstokes$
and $\Gint$ for various values of $T$, linearized 
at EQ1 with spatial resolution $(M_x, M_y, M_z) = (39, 29, 39)$.
Note that these spectra are in general not those of $L+N_U$,
whose eigenvalues determine the stability
of the steady states via the sign of their real parts.
Instead, the $G$'s are operators whose roots are the same as
those of $L+N$ but whose spectra may be quite different.
Newton's method finds the roots of the $G$ operators independent of their stability.
(We do not recommend any of these operators as a means of determining
eigenvalues of $L+N_U$. Other, more precise and more economical methods 
are given in \cite{wang2007lower} via exact diagonalization of the Jacobian, 
in \cite{sanchez2004newton,viswanath2007recurrent,gibson2008visualizing}
via more accurate use of $\Gint$, and
in \cite{barkley1997stokes,IMA,huepe2003stability,tuckerman2015laplacian}
via the inverse Arnoldi method with Stokes preconditioning.)
Our interest is instead in the distribution of the eigenvalues of
the Jacobians $G_U$, which determines the rate of convergence of GMRES in
solving the linear systems \eqref{eq:biglin}.
Broadly speaking, a tightly clustered spectrum
leads to fast convergence, a highly dispersed spectrum to slow convergence.
In order to compare the spectra of $\Gstokes_T$ at different values of $T$,
we have followed \cite{beaume2017adaptive} and rescaled the
eigenvalues $\lambda^\text{Stokes}$ via:
\begin{equation}
  \lambda^{\text{Stokes}} \mapsto \frac{1+T}{T}\lambda^{\text{Stokes}}.
\label{eq:beaumescale}\end{equation}
(Note that scaling does not change the condition number of an operator.)
The scaling factor is $1/T$ in the limit of small $T$ and $1$ for large $T$.
It is designed to remove the $\dt$ ($=T$) dependence from the limiting
behaviors of the
Stokes operator given in~\eqref{eq:smalldt} and~\eqref{eq:largedt}
and thereby facilitate the visual comparison of 
the spectra at different values of $T$ at approximately the same scales.
Using \eqref{eq:beaumescale}, the eigenvalues tend to those of $L+N_U$
as $T\rightarrow 0$ and to those of $-L^{-1}(L+N_U)$ as $T\rightarrow \infty$.
The green points show the spectrum of $\Gstokes$ for $T=10^{-4}$,
a value small enough for $\Gstokes$ to resemble $L+N_U$.
The large negative eigenvalues of the Laplacian (or rather its vector
analogue, the Stokes operator) are prominent outlying features.
Here, the spectrum extends to $-360$, but 
these eigenvalues depend on the spatial resolution, becoming
more negative as the resolution is increased.
It is these eigenvalues that are responsible for the
poor conditioning of $L+N_U$, or equivalently, the stiffness of
the Navier-Stokes equations. 
As $T=\dt$ is increased to $10^{-1}$ (blue points),
the eigenvalues retract towards zero and are contained approximately within
the convex hull of the points $(-11,0)$, $(-1, \pm 13\:i)$, and the origin. 
When $T=\dt$ is increased to 100 (ochre points), 
in the asymptotic regime used in steady-state solving, the spectrum lies 
very close to the line segment between $(-1,-30i)$ and $(-1,30i)$; over 95\% of
the eigenvalues possess real part within $\pm 0.2$ of $-1$. The real range has
decreased drastically, but the imaginary range has increased.
Brynjell-Rahkola et al.~\cite{brynjell2017computing} also show
that the eigenvalues of $\Gstokes_T$ converge towards a curve
as $T$ is increased for the model problem of a Ginzburg-Landau equation.

The eigenvalues of the multi-step operator $\Gint_{T=10}$ 
with $\dt=0.031\,25$ are shown as magenta points.
The Jacobian of $(B_{\dt})^{T/\dt}$ should approach $\exp(T(L+N_U))$
as $\dt \rightarrow 0$.
This leads to the following limits for the eigenvalues $\lambda^\text{int}$ of
$\Gint$, in terms of the eigenvalues $\lambda$ of $L+N_U$:
\begin{equation}
	\lambda^\text{int} \approx  \exp(\lambda T) - 1 \approx
  \left\{\begin{array}{l}\lambda T \\-1 \end{array}\right\}
  \mbox{~~for~~} \left\{\begin{array}{l}|\lambda T| \ll 1
   \\ |\lambda T| \gg 1,  \mbox{~~with~~} \lambda<0 \end{array}\right\}
\label{eq:lims}\end{equation}
which explains why the Jacobian of $\Gint$ typically enjoys superior
conditioning properties.  
This can be seen by examining Fig.~\ref{fig:spectra500}, parts (b) and (c),
which indeed contain values tightly clustered around $-1$ as well as a positive
eigenvalue near 0.6, which is approximately $T=10$ times the
known \cite{wang2007lower} leading eigenvalue 0.06 of EQ1 at $Re=500$.
In contrast, while the ochre points of the $\Gstokes_{100}$ spectrum cluster 
in the real direction, their imaginary parts are spread out
and the operator is therefore more poorly conditioned.

Comparable spectra for $Re=1750$ are shown in Fig.~\ref{fig:spectra1750}.
These differ quantitatively but not qualitatively from those in
Fig.~\ref{fig:spectra500} for $Re=500$.
We use these differences to test the representation of the spectrum of $L+N_U$
by that of a 1D model for an advection-diffusion equation with analytical eigenvalues:
\begin{equation}
  \lambda_k^{\rm model} = -k^2/Re \pm i k U  \label{eq:model}
  \end{equation}
where $k$ represents a wavenumber associated with eigenmodes $e^{ikx}$, whose
images under the diffusion and linearized advection operators, respectively, lead
to the real and imaginary parts of $\lambda_k^{\rm model}$.
(Although $k$ has been used previously to designate a member of the sequence of Krylov vectors,
we use it again here because it is the universal notation for a wavenumber.)
In the following, we assess this model using quantitative comparisons
with the eigenvalues for $\Gstokes_T$, rescaled according to 
Eq.~\eqref{eq:beaumescale}, and $\Gint_{10}$.
We first consider the most negative real eigenvalue of $L+N_U$, which is
predicted by \eqref{eq:model} to be proportional to $1/Re$.
The most negative eigenvalue among the green points of $\Gstokes_{0.0001}$
in Fig.~\ref{fig:spectra1750}(a) is $-105$, compared to $-360$ for $Re=500$.
The ratio of eigenvalues $360/105$ is $3.4$, almost exactly the inverse of the ratio
of Reynolds numbers $1750/500=3.5$, as predicted by~\eqref{eq:model}.
Our second test concerns the imaginary part of the spectrum of $\Gstokes_T$. 
For $T$ large, 
the model spectrum \eqref{eq:model} is transformed under Stokes preconditioning
to the vertical line
\begin{equation}
\lambda_k^{\rm Stokes} \approx \frac{\lambda_k^{\rm model}}{k^2/Re} = -1 \pm i U Re / k ,
\end{equation}
which resembles the spectrum of $\Gstokes_{100}$ aligned along the imaginary axis. 
To verify the scaling, note that for $Re=1750$, the largest imaginary part among
the ochre points of $\Gstokes_{100}$ is $\pm 100$, compared with $\pm 35$ for $Re=500$.
The ratio between these values is 2.9, close to the ratio between the
Reynolds numbers.

For our third test, we use \eqref{eq:model} to estimate the radius $r$ of the circle
surrounding $(-1,0)$ which contains 95\% of the eigenvalues of $\Gint$.
Using the limits in~\eqref{eq:lims}, we assume that 
these eigenvalues correspond to 
those of \eqref{eq:model} with $k>k_\ast$. We have
\begin{align}
\left|\lambda_k^\text{int} + 1\right| &\lesssim e^{-T k_\ast^2/Re} = r 
\intertext{and therefore}
T k_\ast^2/Re &= \ln(1/r), \label{eq:TRe}
\intertext{yielding another expression for the ratio between Reynolds numbers $Re_1$ and
$Re_2$:}
\frac{Re_2}{Re_1} &= \frac {\ln(1/r_1)}{\ln(1/r_2)}. \label{eq:Redep}
\intertext{The radii measured for $Re_1=500$ and $Re_2=1750$ are $r_1=0.000\,84$ and
$r_2=0.153$, respectively, leading to}
\frac{\ln(1/0.000\,84)}{\ln(1/0.153)} &= \frac{7.08}{1.877}=3.77,
\end{align}
which is again very close to $1750/500=3.5$.

When $T$ is changed and $Re$ kept constant,
\eqref{eq:TRe} implies that the 
left-hand-side of \eqref{eq:Redep} should be replaced by $T_1/T_2$.
We have verified this for $Re=500$ by computing the radius $0.049$
containing 95\% of the eigenvalues for $\Gint_{T=1}$.
Comparing this to the radius corresponding to $\Gint_{T=10}$ yields
\begin{equation}
  \frac{\ln(1/0.00084)}{\ln(1/0.049)}=10.003
\end{equation}
very close to the ratio of the $T$ values.
To preserve the conditioning of $\Gint$ as $Re$ is increased,
the ratio $T/Re$ should be kept constant. Since the time taken
to act with $\Gint$ is proportional to $T/\dt$, we arrive at the useful
result that the CPU time taken by $\Gint$ to calculate steady states
or traveling waves is, in the absence of changes in the spatial
or temporal resolution, proportional to the Reynolds number. 
The number of timesteps taken by $\Gstokes$ is also approximately 
proportional to $Re$, as shown in Fig.~\ref{fig:M1L}(left)
and in \cite{brynjell2017computing}.

The true spectrum of $L+N_U$ is certainly more complicated than \eqref{eq:model}.
The green dots in Fig.~\ref{fig:spectra500} and \ref{fig:spectra1750}
show many features not present in \eqref{eq:model} and indeed,
the eigenvalues of $L$ and $N_U$ are not as simple as $-k^2/Re$ and $\pm i k U$.
Note, however, that the only features of \eqref{eq:model} that we have used are
shared by the more general model
\begin{equation}
  {\lambda_k^{\rm model}}^{~\prime}= -f(k)/Re \pm i g(k) \label{eq:modelgen}
  \end{equation}
in which $k$ serves merely as an index
enumerating the eigenvalues. The astonishing accuracy of
the predictions of this model leads us to believe that it provides a
good representation of the basic shape of the actual spectra and operators.
Moreover, it provides a simple explanation of the dependence of the spectra
on $Re$ and $T$ and by extension, the conditioning properties of the operators. 

\subsection{Challenges}

We close by mentioning the limitations and applicability 
of the Stokes preconditioning method.
First, the single-step Stokes operator can be used only for computing steady
states and traveling waves and not for calculating general periodic orbits
or more complex dynamical states. 
It seems unlikely that any generalization would be possible, 
since even describing or defining these 
objects necessarily requires multiple time steps.
However, Stokes preconditioning can be used to rapidly compute leading 
eigenvalues \cite{barkley1997stokes,IMA,huepe2003stability,tuckerman2015laplacian}
and optimal forcing \cite{brynjell2017computing} via the inverse power or Arnoldi method.
Second, Stokes preconditioning is based on the utility 
and speed of inverting elliptic (Laplacian, Stokes, or Helmholtz) operators.
Concerning utility, Stokes preconditioning
is an effective preconditioner for the Jacobian $L+N_U$ under precisely the
same conditions that implicit timestepping of the diffusive/viscous operator $L$
is effective.
Concerning speed, the most favorable conditions occur when
the elliptic operators are inverted directly rather than iteratively, 
which is easy in a tensor-product domain
\cite{tuckerman1989steady,lynch1964direct,haidvogel1979accurate,vitoshkin2013direct}.
Codes such as Nek5000 \cite{Nekton}, which use spectral elements to represent
flexible user-defined geometries, typically do not do so.
Instead, these codes use iterative methods whose convergence relies on the Helmholtz
operators $I-\dt L$ being close to the identity, which is not the case
when $\dt$ is taken large. Similarly, codes sometimes impose constraints, notably
incompressibiity, to a power of $\dt$ rather to machine accuracy;
this too is not compatible with large $\dt$.
However, based on the experience of previous researchers using
spectral element methods
we believe that these objections can be overcome.

We have seen that Stokes preconditioning is extremely economical
at the lower Reynolds number ranges of our investigation,
achieving a 50-fold economy in CPU time.
This factor, however, decreases as the Reynolds number increases,
and we have been able to understand the reason for this.
As $Re$ increases, the Jacobian $L+N_U$ deviates increasingly from $L$, 
preconditioning by $L^{-1}$ becomes less effective, and so 
the number $K$ of iterations necessary to solve each linear system increases.
The consequences of this are particularly severe when GMRES is used,
since part of its algorithm requires a time which is quadratic in $K$.
We can propose several remedies for this, in increasing order of
difficulty and effectiveness.
First, $K$ may be reduced by relaxing the convergence criterion
for solving the linear equation \eqref{eq:biglin}.
This has the counterbalancing effect of increasing the number of
Newton steps required, but some improvement is possible.
A more promising approach is to use a different method for solving the linear systems.
Conjugate-gradient-type methods other than GMRES, notably BiCGSTAB,
do not retain all $K$ Krylov vectors throughout the calculation.
Hence, they do not store $K$ velocity fields 
and no portion of their operation count scales like $K^2$.
Although BiCGSTAB has been widely used in the applications of
Stokes preconditioning detailed in the introduction,
we have not succeeded in doing so in Channelflow or in Openpipeflow.
The reason for this is probably given in Figs.~\ref{fig:spectra500} and
\ref{fig:spectra1750}: 
as the Reynolds number is increased, the eigenvalues associated with $\Gstokes$
acquire large imaginary parts and
BiCGSTAB is known to converge badly in this case
\cite{gutknecht1993variants,sleijpen1993bicgstab}.
Related methods, such as BiCGSTAB($\ell$) and IDR \cite{sonneveld2008idr} 
that act better with such matrices have been proposed.
The most promising idea, inspired by Figs.~\ref{fig:spectra500} and
\ref{fig:spectra1750}
and the model \eqref{eq:modelgen} is to incorporate linear portions of
$N$ into $L$, in particular the linearization around the laminar flow
\begin{equation}
\left(\bU_{\rm lam}\cdot \nabla\right)\bu + \left(\bu \cdot \nabla\right)\bU_{\rm lam}
\end{equation}
so as to include a large part of the advective term in the preconditioning.

\begin{acknowledgement}
We thank Dwight Barkley and John Gibson for their contributions.
We acknowledge the support of TRANSFLOW, provided by the Agence Nationale de
la Recherche (ANR).

\end{acknowledgement}
\clearpage
\section*{Appendix}
\begin{figure}
\centerline{\includegraphics[width=\textwidth]{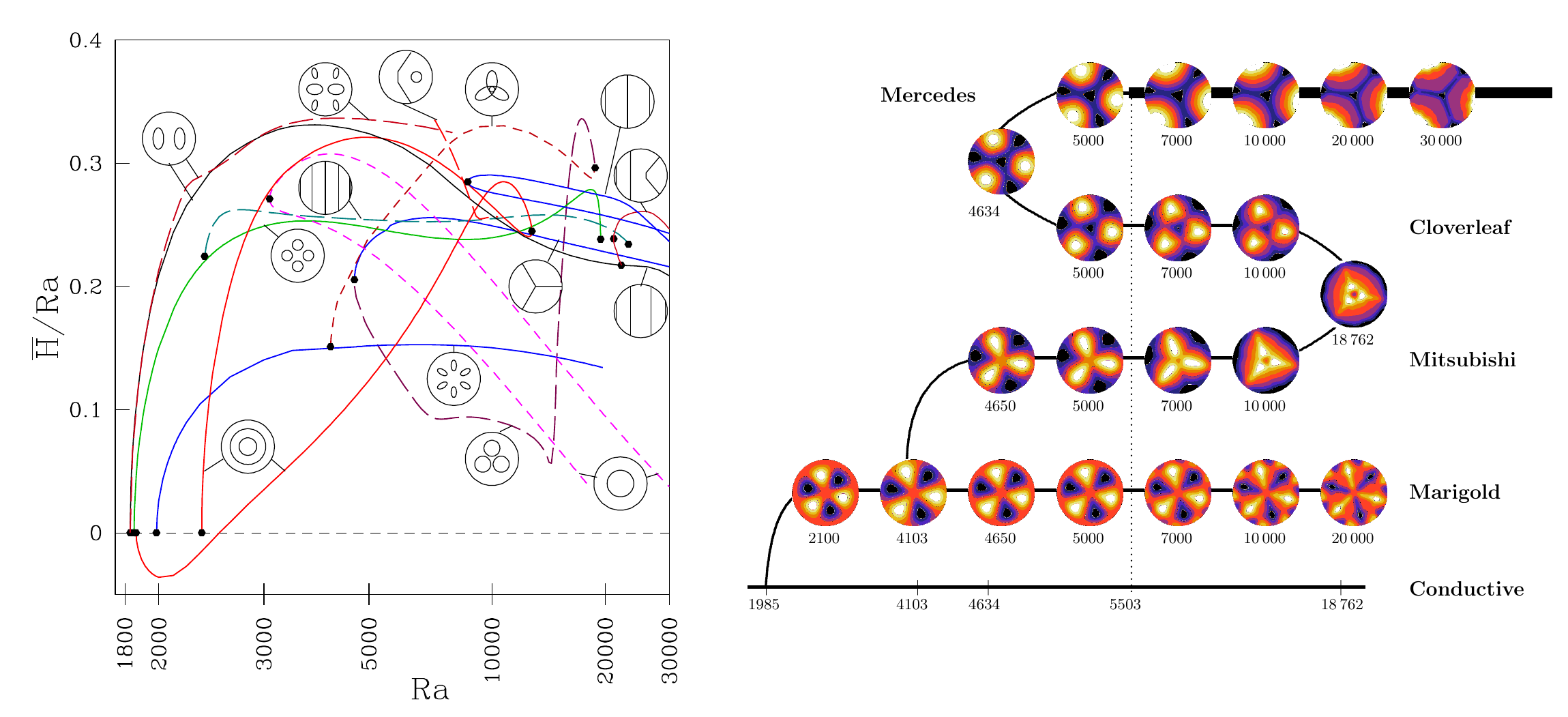}}
\caption{Rayleigh-B\'enard convection in a cylinder of aspect ratio
	with radius / height $=2$, $Pr=6.7$ and insulating lateral boundaries.
  Left: Bifurcation diagram shows 17 branches of steady states,
  with azimuthal symmetries $m=2$ (pizza, four-roll),
  $m=0$ (two-tori, torus), $m=3$ (marigold, Mitsubishi,
  cloverleaf, Mercedes), $m=1$ (dipole, three-roll, tiger,
  asymmetric three-roll). Right: Partial schematic diagram showing 
  branches with $m=3$ symmetry. Transition from conductive state to
  marigold and then Mitsubishi branches occur via circle and ordinary
  pitchfork bifurcations, respectively,
  and to cloverleaf and Mercedes branches via two successive
  saddle-node bifurcations. The only stable states are on 
  a portion of the Mercedes branch, shown by the thick curve.
  The results are from a pseudospectral simulation with
  $(M_r,M_\theta,M_z)=(60,130,30)$.
  From Boro\'nska \& Tuckerman \cite{boronska2010extreme}.}
\label{fig:boronska}
\centerline{\includegraphics[width=\textwidth]{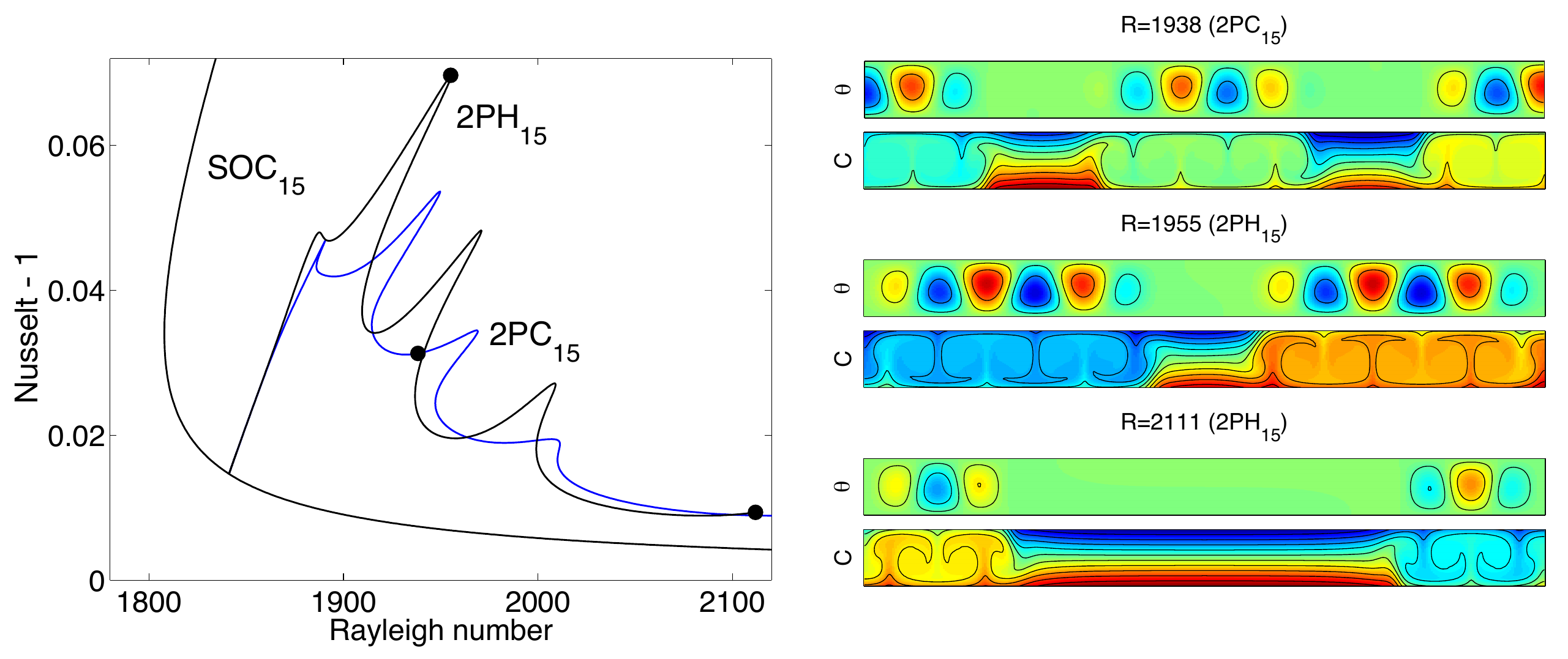}}
\caption{Binary fluid convection in a domain with aspect ratio
  width / height $= 14$, Neumann boundary conditions, Prandtl and Lewis
  numbers $Pr=7$, $Le=0.01$ and separation ratio $S=-0.1$.  Left: Partial
  bifurcation diagram showing two-pulse point-symmetric states based
  on 15 rolls. Right: Temperature and concentration fields for 
  the three solutions indicated as dots on the bifurcation diagram.
From Mercader, Batiste, Alonso \& Knobloch \cite{mercader2011convectons}.}
\label{fig:mercader}
\end{figure}

\begin{figure}
\centerline{
\includegraphics[width=0.8\textwidth]{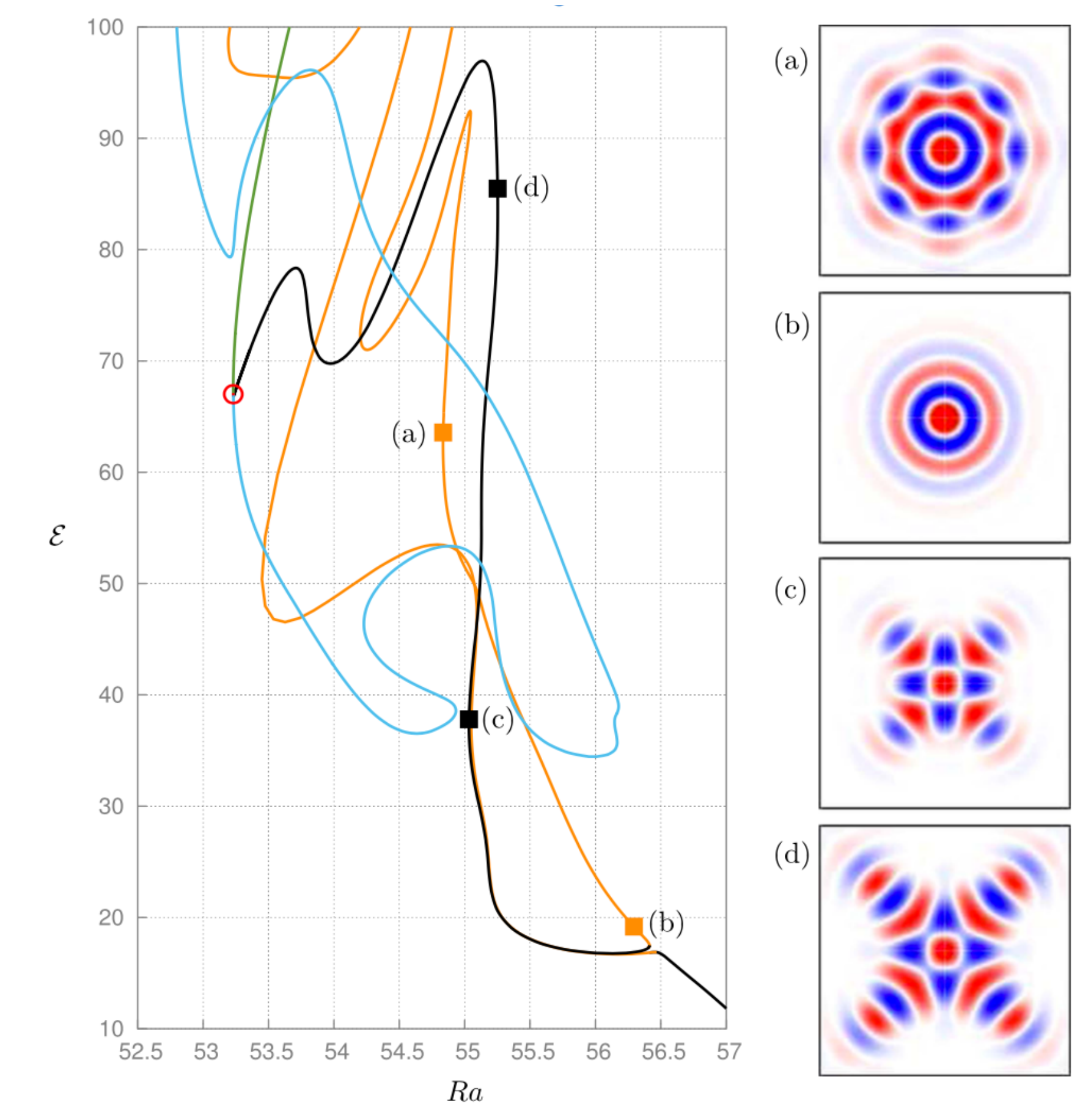}}
\caption{Three-dimensional binary fluid convection in a porous medium
  of size $6\times 6 \times 1$.
  Left: bifurcation diagram. Right: vertical velocity at mid-layer.
The transition from a four-armed structure with arms oriented along the
diagonals (panels (c) and (d), black curve) to an eight-armed
structure with arms oriented along both the diagonals and the principal axes of the domain (panel (a), orange curve) via a target pattern
(panel (b)). Simulations use a spectral element method with 6 elements
in the quarter domain, each with $(23,23,17)$ points.
From LoJacono, Bergeon \& Knobloch \cite{jacono2017complex}.}
\label{fig:LoBeKn}
\end{figure}

\begin{figure}
\includegraphics[width=\textwidth]{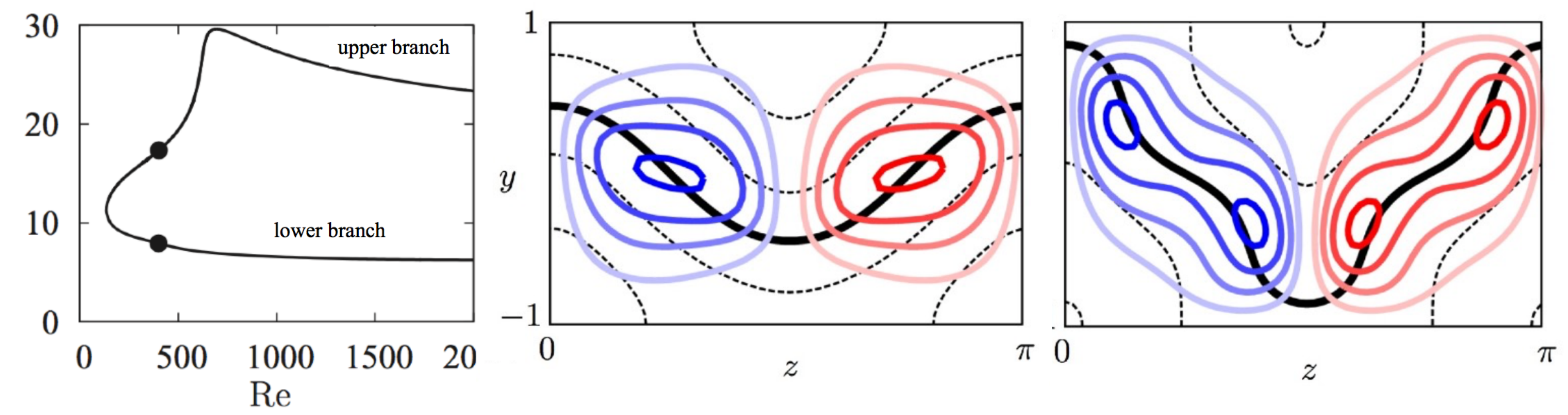}
\caption{Steady states of a streamwise-independent reduced model
  for plane Couette flow. Bifurcation diagram on left.
Representative states from lower 
(middle) and upper (right) branches at $Re\approx 1000$.
  Colored contours show the streamfunction of streamwise rolls,
  while the black curves show contours of the streamwise velocity.
From Beaume, Chini, Julien \& Knobloch \cite{beaume2015reduced,beaume2017adaptive}.}
\label{fig:Beaume}
\end{figure}

\addcontentsline{toc}{section}{Appendix}
This Appendix presents samples in Figs. \ref{fig:boronska},
\ref{fig:mercader}, \ref{fig:LoBeKn} and \ref{fig:Beaume}
of previous computations carried out by Stokes preconditioning. 

\end{document}